\pdfoutput=1
\newcommand*{\ATLASLATEXPATH}{latex/}
\documentclass[cernpreprint,texlive=2016,txfonts,UKenglish]{latex/atlasdoc}

\usepackage[subcaption]{\ATLASLATEXPATH atlaspackage}
\usepackage{\ATLASLATEXPATH atlasbiblatex}

\usepackage{\ATLASLATEXPATH atlascontribute}

\usepackage[BSM]{\ATLASLATEXPATH atlasphysics}

\graphicspath{{logos/}{figures/}}

\usepackage{PUBstopZH-defs}

\AtlasTitle{Search for direct top squark pair production in events with a Higgs or $Z$ boson, and missing transverse momentum in $\sqrt{s}=13$ TeV $pp$ collisions with the ATLAS detector}

\author{The ATLAS Collaboration}

\PreprintIdNumber{CERN-EP-2017-106}

\AtlasJournal{JHEP}

\AtlasAbstract{%
A search for direct top squark pair production resulting in events with either a same-flavour opposite-sign dilepton pair with invariant mass compatible with a $Z$ boson or a pair of jets compatible with a Standard Model (SM) Higgs boson ($h$) is presented. 
Requirements on additional leptons, jets, jets identified as originating from $b$-quarks, and missing transverse momentum are imposed to target the other decay products of the top squark pair.
The analysis is performed using proton-proton collision data at $\sqrt{s}=13$ TeV collected with the ATLAS detector at the LHC in 2015--2016, corresponding to an integrated luminosity of 36.1 fb$^{-1}$. No excess is observed in the data with respect to the SM predictions.
The results are interpreted in two sets of models. In the first set, direct production of pairs of lighter top squarks ($\tilde{t}_1$) with long decay chains involving $Z$ or Higgs bosons is considered. The second set includes direct pair production of the heavier top squark pairs ($\tilde{t}_2$) decaying via $\tilde{t}_2 \rightarrow Z \tilde{t}_1 $ or $\tilde{t}_2 \rightarrow h \tilde{t}_1$. The results exclude at 95\% confidence level $\tilde{t}_2$ and $\tilde{t}_1$ masses up to about 800 GeV, extending the exclusion region of supersymmetric parameter space covered by previous LHC searches.
}

\hypersetup{pdftitle={ATLAS document},pdfauthor={The ATLAS Collaboration}}

\begin{document}

\maketitle



\section{Introduction}
\label{sec:intro}

Supersymmetry (SUSY)~\cite{Golfand:1971iw,Volkov:1973ix,Wess:1974tw,Wess:1974jb,Ferrara:1974pu,Salam:1974ig} 
is one of the most studied extensions of the Standard Model (SM). It predicts new bosonic partners for the existing 
fermions and fermionic partners for the known bosons.
If $R$-parity is conserved~\cite{Farrar:1978xj}, SUSY particles are produced in pairs and the lightest supersymmetric 
particle (LSP) is stable, providing a possible dark-matter candidate. The SUSY partners of the charged (neutral) Higgs bosons and electroweak gauge bosons mix to form
the mass eigenstates known as charginos ($\tilde{\chi}^{\pm}_{k}$, $k = 1, 2$) and neutralinos ($\tilde{\chi}^{0}_{m}$, $m = 1, \ldots, 4$), where the increasing index denotes increasing mass.
The scalar partners of right-handed and left-handed quarks, \squarkR and \squarkL, mix to form two mass eigenstates, $\tilde{q}_{1}$ and $\tilde{q}_{2}$, with $\tilde{q}_{1}$ defined to be the lighter of the two. 
To address the SM hierarchy problem~\cite{Sakai:1981gr,Dimopoulos:1981yj,Ibanez:1981yh,Dimopoulos:1981zb}, \TeV-scale masses are required~\cite{Barbieri:1987fn,deCarlos:1993yy} for the supersymmetric partners of the gluons (gluinos, ~\gluino) and the top squarks~\cite{Inoue:1982pi,Ellis:1983ed}.
 Furthermore, the higgsino is required not to be heavier than a few hundred \GeV.

Top squark production with Higgs ($h$) or $Z$ bosons in the decay chain can appear either in production of the lighter top squark mass eigenstate (\stopone) decaying via $\stopone\to t\ninotwo$ with $\ninotwo\to h/Z\ninoone$, 
or in production of the heavier top squark 
mass eigenstate (\stoptwo) decaying via $\stoptwo\to h/Z \stopone$, as illustrated in Figure~\ref{fig:susyfeynman}. 
Such signals can be discriminated from the SM top quark pair production ($\ttbar$) background by requiring a pair of $b$-tagged jets originating from the $h\to\bbbar$ decay or a same-flavour opposite-sign lepton pair originating from the $Z\to\ell^+\ell^-$ decay.
Although the pair production of \stopone has a cross-section larger than that of the \stoptwo, and their decay properties
can be similar, searches for the latter can provide additional sensitivity in regions where the \stopone falls in a phase space  difficult to experimentally discriminate from the background due to the similarities in kinematics with $\ttbar$ pair production,
such as scenarios
where the lighter top squark is only slightly heavier than the sum of the masses of the top quark
and the lightest neutralino (\ninoone). 
 
Simplified models~\cite{Alwall:2008ve,Alwall:2008ag,Alves:2011wf} are used for the analysis optimisation and interpretation of the results. 
In these models, direct top squark pair production is considered and all SUSY particles are decoupled except for the top squarks and the neutralinos involved in their decay. In all cases the $\ninoone$ is assumed to be the LSP. 
Simplified models featuring direct $\stopone$ production with $\stopone \to t\ninotwo$ and either $\ninotwo \to Z\ninoone$ or $\ninotwo \to h\ninoone$ are considered. 
Additional simplified models featuring direct $\stoptwo$ production with $\stoptwo\to Z\stopone$ or $\stoptwo\to h\stopone$ decays and $\stopone \to t\ninoone$ are also considered, where the mass difference between the lighter top squark and the neutralino is set to 180~\GeV, a region of the mass parameter space not excluded by previous searches for \stopone with mass greater than 191~\GeV~\cite{AtlasSummary}.

This paper presents the results of a search for top squarks in final states with $h$ or $Z$ bosons at $\sqrt{s}=13$~\TeV\ using the data collected by the ATLAS experiment~\cite{PERF-2007-01} in proton--proton ($pp$) collisions during 2015 and 2016, corresponding to 36.1~\ifb. Searches for direct \stopone pair production have been performed by the ATLAS Collaboration at $\sqrt{s}=7,8$~\TeV\ using LHC Run-1 data~\cite{AtlasSummary,SUSY-2012-03} and $\sqrt{s}=13~\TeV$ with 2015 data~\cite{SUSY-2015-02} and by the CMS Collaboration at $\sqrt{s}=8$~\TeV~\cite{CMS-SUS-13-011,CMS-SUS-13-004,CMS-SUS-13-019,CMS-SUS-14-021,CMS-SUS-14-015,CMS-SUS-13-023}, searches for direct \stoptwo\ production were performed at $\sqrt{s}=8$~\TeV\ by both collaborations~\cite{AtlasSummary,Stop2_Run1,Stop2_Run1cms}.

\begin{figure}[htb]
\begin{center}
    \begin{subfigure}[t]{0.5\textwidth}
        \centering
                \includegraphics[width=0.8\textwidth]{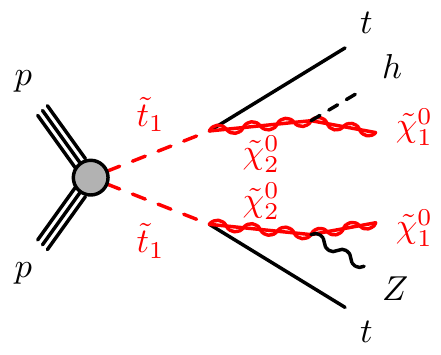}
                \caption{}
        \end{subfigure}%
    \begin{subfigure}[t]{0.5\textwidth}
        \centering
        \includegraphics[width=0.8\textwidth]{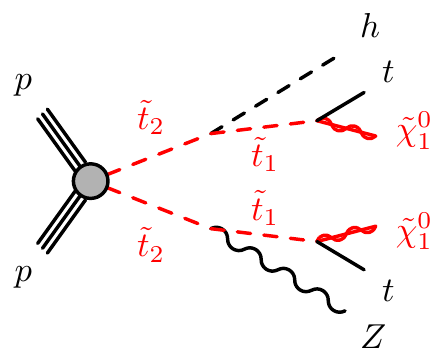}
        \caption{}
        \end{subfigure}%
\end{center}
\caption{Diagrams for the top squark pair production processes considered in this analysis: (a) $\stopone \to t \ninotwo$ and $\ninotwo\to h/Z\ninoone$ decays, and (b) $\stoptwo \to h/Z \stopone$ and $\stopone\to t\ninoone$ decays.}
\label{fig:susyfeynman}
\end{figure}

\section{ATLAS detector}
\label{sec:detector}
%
The ATLAS experiment~\cite{PERF-2007-01} is a multi-purpose particle detector with a forward-backward symmetric cylindrical
geometry and nearly $4\pi$ coverage in solid angle.\footnote{ATLAS uses
  a right-handed coordinate system with its origin at the nominal
  interaction point (IP) in the centre of the detector and the
  $z$-axis along the beam pipe. The $x$-axis points from the IP to the
  centre of the LHC ring, and the $y$-axis points upward. Cylindrical
  coordinates ($r$, $\phi$) are used in the transverse plane, $\phi$
  being the azimuthal angle around the beam pipe. The pseudorapidity
  is defined in terms of the polar angle $\theta$ as $\eta = -\ln
  \tan(\theta/2)$. Rapidity is defined as $y=0.5 \ln\left[(E + p_z )/(E - p_z )\right]$ where $E$ denotes the energy and $p_z$ is the component of the momentum along the beam direction.}
The interaction point is surrounded by an inner detector (ID) for tracking, a
calorimeter system, and a muon spectrometer.

The ID provides precision tracking of charged particles for
pseudorapidities $|\eta| < 2.5$ and is surrounded by a superconducting solenoid providing a \SI{2}{T} axial magnetic field.
It consists of silicon pixel and microstrip detectors inside a
transition radiation tracker.  
One significant upgrade for the running period at $\sqrt{s}=13$~\TeV\ is the presence of the
insertable B-layer~\cite{CERN-LHCC-2010-013}, an additional pixel layer close to the interaction point, which 
provides high-resolution hits at small radius to improve the tracking performance.

In the pseudorapidity region $|\eta| < 3.2$, high-granularity lead/liquid-argon (LAr)
electromagnetic (EM) sampling calorimeters are used.
A steel/scintillator tile calorimeter measures hadron energies for
$|\eta| < 1.7$.
The endcap and forward regions, spanning $1.5<|\eta| <4.9$, are 
instrumented with LAr calorimeters for both the EM and hadronic energy
measurements.

The muon spectrometer consists of three large superconducting toroids
with eight coils each, and a system of trigger and precision-tracking chambers, 
which provide triggering and tracking capabilities in the
ranges $|\eta| < 2.4$ and $|\eta| < 2.7$, respectively.

A two-level trigger system is used to select events~\cite{Aaboud:2016leb}. The first-level
trigger is implemented in hardware and uses a subset of the detector
information. This is followed by the software-based high-level trigger stage,
which runs offline reconstruction and calibration software, reducing the event rate to about \SI{1}{kHz}.

\section{Data set and simulated event samples}
\label{sec:dataMC}

The data were collected by the ATLAS detector during 2015 with a peak 
instantaneous luminosity of ${\cal L}=5.2\times 10^{33}$~cm$^{-2}$s$^{-1}$, 
and during 2016 with a peak instantaneous luminosity of ${\cal L}=1.4\times 10^{34}$~cm$^{-2}$s$^{-1}$,
resulting in a mean number of additional $pp$ interactions per bunch crossing 
(pile-up) of $\langle \mu \rangle = 14$ in 2015 and $\langle \mu \rangle = 24$ in 2016.
Data quality requirements are applied to ensure that all subdetectors were
operating at nominal  
conditions, and that LHC beams were in stable-collision mode. The integrated
luminosity of the resulting  
data set is 36.1~\ifb\ with an uncertainty of $\pm 3.2\%$. 
The luminosity and its uncertainty are derived following a methodology similar to that detailed in
Ref.~\cite{DAPR-2011-01} from a preliminary calibration of the
luminosity scale using a pair of $x$--$y$ beam-separation scans performed in
August 2015 and May 2016.

Monte Carlo (MC) simulated event samples are used to aid in the
estimation of the background from SM processes and to model the SUSY signal. 
The choices of MC event generator, parton shower and hadronisation, the cross-section normalisation, the parton distribution function (PDF) set and the set of tuned parameters (tune) for the underlying event of these samples are summarised in Table~\ref{tab:MC}, and more details of the event generator configurations can be found in Refs.~\cite{ATL-PHYS-PUB-2016-004,ATL-PHYS-PUB-2016-003,ATL-PHYS-PUB-2016-002,ATL-PHYS-PUB-2016-005}.
Cross-sections calculated at next-to-next-to-leading order (NNLO) in quantum chromodynamics (QCD) including resummation of next-to-next-to-leading logarithmic (NNLL) soft-gluon terms are used for top quark production processes. 
For production of top quark pairs in association with vector and Higgs bosons, cross-sections calculated at next-to-leading order (NLO) are used, and the event generator cross-sections from \SHERPA (at NLO for most of the processes) are used when normalising the multi-boson backgrounds.
In all MC samples, except those produced by \SHERPA, the {\textsc EvtGen}~v1.2.0 program~\cite{EvtGen} is used to model the properties of the bottom and charm hadron decays.

\begin{table*}[t]
\caption{Simulated signal and background event samples: the corresponding event generator, the parton shower, the cross-section normalisation, the PDF set and the underlying-event tune are shown.}
\label{tab:MC}
\begin{center}
\resizebox{\textwidth}{!}
{\small
\begin{tabular}{l l c c c c}
\hline
Physics process    & Generator & Parton shower & Cross-section & PDF set & Tune \\
                   &	      & 	      & normalisation & 	&      \\
\hline
SUSY Signals    	   & \AMCATNLO 2.2.3~\cite{Alwall:2014hca} & \PYTHIA 8.186~\cite{Sjostrand:2007gs} & NLO+NLL~\cite{Beenakker:1996ch,Kulesza:2008jb,Kulesza:2009kq,Beenakker:2009ha,Beenakker:2011fu} & NNPDF2.3LO~\cite{Lai:2010vv} & A14~\cite{pub-2014-021} \\
\hline
  $W(\rightarrow \ell\nu)$ + jets              & \SHERPA 2.2.1~\cite{Gleisberg:2008ta}    & \SHERPA 2.2.1     & NNLO~\cite{Catani:2009sm}  & NLO CT10~\cite{Lai:2010vv}    & \SHERPA default \\
  $Z/\gamma^{*}(\rightarrow \ell \ell)$ + jets & \SHERPA 2.2.1   & \SHERPA 2.2.1       & NNLO~\cite{Catani:2009sm}  & NLO CT10         & \SHERPA default\\ \hline
  
  $t\bar{t}$              & {\textsc powheg-box} v2~\cite{Alioli:2010xd} & \PYTHIA6.428~\cite{Sjostrand:2006za}  & NNLO+NNLL~\cite{Czakon:2013goa,Czakon:2012pz,Czakon:2012zr,Baernreuther:2012ws,Cacciari:2011hy,Czakon:2011xx}  & NLO CT10  &\textsc{Perugia2012}~\cite{Skands:2010ak}\\
  
  Single-top              &&&&\\
  ($t$-channel)           & {\textsc powheg-box} v1 & \PYTHIA6.428 & NNLO+NNLL~\cite{Kidonakis:2011wy}  & NLO CT10f4   & \textsc{Perugia2012}\\
  Single-top              &&&&\\
  ($s$- and $Wt$-channel) & {\textsc powheg-box} v2 & \PYTHIA6.428  & NNLO+NNLL~\cite{Kidonakis:2010ux,Kidonakis:2010tc}  & NLO CT10   & \textsc{Perugia2012}\\ \hline

$t\bar{t} W/Z/\gamma^{*}$ & \AMCATNLO 2.2.2   & \PYTHIA 8.186	& NLO~\cite{Alwall:2014hca}   & NNPDF2.3LO & A14    \\
Diboson    	   & \SHERPA 2.2.1       & \SHERPA 2.2.1	& Generator NLO	 & CT10	     & \SHERPA default \\
\hline
$t\bar{t}h$	   & \AMCATNLO 2.2.2        & \HERWIG 2.7.1~\cite{Corcella:2000bw}          & NLO~\cite{Dittmaier:2012vm}   & CTEQ6L1 & A14  \\
$Wh$, $Zh$     & \AMCATNLO 2.2.2       & \PYTHIA 8.186   & NLO~\cite{Dittmaier:2012vm}   & NNPDF2.3LO     & A14  \\
$t\bar{t} WW$, $\ttbar\ttbar$  & \AMCATNLO 2.2.2       & \PYTHIA 8.186   & NLO~\cite{Alwall:2014hca}	 & NNPDF2.3LO	  & A14  \\
$tZ$, $tWZ$, $t\bar{t}t$    & \AMCATNLO 2.2.2       & \PYTHIA 8.186   & LO                   	    & NNPDF2.3LO     & A14  \\
Triboson	   & \SHERPA 2.2.1        & \SHERPA 2.2.1    & Generator LO, NLO            & CT10	     & \SHERPA default \\
\hline
\end{tabular}
}
\end{center}
\end{table*}

SUSY signal samples are generated from leading-order (LO) matrix elements with up to two extra partons, 
using the \AMCATNLO v2.2.3 event generator interfaced to \PYTHIA 8.186 with the {\textsc A14} tune for the modelling of the SUSY decay chain, 
parton showering, hadronisation and the description of the underlying event. 
Parton luminosities are provided by the {\textsc NNPDF23LO} PDF set. 
Jet--parton matching is realised following the CKKW-L prescription~\cite{Lonnblad:2011xx}, with a matching scale set to one quarter of the pair-produced superpartner mass. In all cases, the mass of the top quark is fixed at 172.5~\GeV.  
Signal cross-sections are calculated to NLO in the strong coupling constant, adding the resummation of soft-gluon emission at next-to-leading-logarithmic accuracy (NLO+NLL)~\cite{Beenakker:1997ut,Beenakker:2010nq,Beenakker:2011fu}. The nominal cross-section and the uncertainty are based on 
predictions using different PDF sets and factorisation and renormalisation scales, as described in Ref.~\cite{Borschensky:2014cia}. 
 
To simulate the effects of additional $pp$ collisions in the same and nearby bunch crossings, additional interactions are generated using the soft QCD processes as provided by \PYTHIA 8.186 
with the A2 tune~\cite{ATLAS-PHYS-PUB-2012-003} and the MSTW2008LO PDF set~\cite{Martin:2009iq}, and overlaid onto each simulated hard-scatter event. 
The MC samples are reweighted so that the pile-up distribution matches the one observed in the data. The MC samples are processed through an ATLAS detector simulation~\cite{Aad:2010ah} based on {\textsc Geant4}~\cite{Agostinelli:2002hh} or, in the case of $\ttbar t$ and the SUSY signal samples, a fast simulation using a parameterisation of the calorimeter response and {\textsc Geant4} for the other parts of the detector~\cite{ATL-PHYS-PUB-2010-013}.
All MC samples are reconstructed in the same manner as the data. 
 
\section{Event selection}
\label{sec:selection}

Candidate events are required to have a reconstructed vertex~\cite{ATL-PHYS-PUB-2015-026} with at least two associated tracks with transverse momentum (\pT) larger than 400~\MeV\ which are consistent with originating from the beam collision region in the $x$--$y$ plane. The vertex with the highest scalar sum of the squared transverse momentum of the associated tracks is considered to be the primary vertex of the event. 

Two categories of leptons (electrons and muons) are defined: ``candidate'' and ``signal'' (the latter being a subset of the ``candidate'' leptons satisfying tighter selection criteria). 
Electron candidates are reconstructed from isolated electromagnetic calorimeter energy deposits matched to ID tracks and are required to have $|\eta|<2.47$, a transverse momentum $\pT>\SI{10}{\GeV}$, and to pass a ``loose'' likelihood-based identification requirement~\cite{Aaboud:2016vfy,ATL-PHYS-PUB-2015-041}. 
The likelihood input variables include measurements of shower shapes in the calorimeter and track properties in the ID. 

Muon candidates are reconstructed in the region $|\eta|<2.5$ from muon spectrometer tracks matching ID tracks.
Candidate muons must have $\pT>\SI{10}{\GeV}$ and pass the medium identification requirements defined in Ref.~\cite{PERF-2015-10}, based on the number of hits in the different ID and muon spectrometer subsystems, and on the significance of the charge to momentum ratio $q/p$.

Jets are reconstructed from three-dimensional energy clusters in the calorimeter~\cite{ATL-PHYS-PUB-2015-036} using the anti-$k_t$ jet clustering algorithm~\cite{Cacciari:2008} with a radius parameter $R=0.4$. Only jet candidates with $\pT>\SI{30}{\GeV}$ and $|\eta|<2.5$ are considered as selected jets in the analysis. Jets are calibrated as described in Refs.~\cite{ATLAS-PHYS-PUB-2015-015,ATLAS-CONF-2015-002}, and the expected average energy contribution from pile-up clusters is subtracted according to the jet area~\cite{ATLAS-PHYS-PUB-2015-015}.
In order to reduce the effects of pile-up, for jets with $\pt<\SI{60}{\GeV}$ and $|\eta|<2.4$ a significant fraction of the tracks associated with each jet must have an origin compatible with the primary vertex, as defined by the jet vertex tagger~\cite{PERF-2014-03}. 

Events are discarded if they contain any jet with $\pT>\SI{20}{\GeV}$ not satisfying basic quality selection criteria designed to reject detector noise and non-collision backgrounds~\cite{ATLAS-CONF-2010-038}.

Identification of jets containing $b$-hadrons is performed with 
a multivariate discriminant that makes use of track impact parameters and reconstructed secondary vertices ($b$-tagging)~\cite{PERF-2012-04,ATL-PHYS-PUB-2016-012}.
A requirement is chosen corresponding to a 77\% average efficiency obtained for $b$-quark jets in simulated $\ttbar$ events. 
The rejection factors for light-quark and gluon jets, $c$-quark jets and $\tau \rightarrow \mathrm{hadrons} + \nu$ decays in simulated $\ttbar$ events are approximately 380, 12 and 54, respectively.
To compensate for differences between data and MC simulation in the $b$-tagging efficiencies and mis-tag rates, 
correction factors are applied to the simulated samples~\cite{ATL-PHYS-PUB-2016-012}. 

Jet candidates within $\Delta R =\sqrt{(\Delta y)^2+(\Delta\phi)^2} = 0.2$ of a lepton candidate are discarded, unless the jet has a value of the $b$-tagging discriminant larger than the value corresponding to approximately 85\% $b$-tagging efficiency, in which case the lepton is discarded since it probably originated from a semileptonic $b$-hadron decay. 
Any remaining electron candidate within $\Delta R=0.4$ of a non-pile-up jet, and any muon candidate within $\Delta R= \operatorname{min}\{0.4, 0.04 + \pt(\mu)/\SI{10}{\GeV}\}$ of a non-pile-up jet is discarded. 
In the latter case, if the jet has fewer than three associated tracks or the muon \pt is larger than half of the jet $\pt$, the muon is retained and the jet is discarded instead 
to avoid inefficiencies for high-energy muons undergoing significant energy loss in the calorimeter. 
Any muon candidate reconstructed with ID and calorimeter information only which shares an ID track with an electron candidate is removed. Finally, any electron candidate sharing an ID track with a remaining muon candidate is also removed.

Tighter requirements on the lepton candidates are imposed, which are then referred to as ``signal'' electrons or muons.
Signal electrons must satisfy the ``medium'' likelihood-based identification requirement as defined in Ref.~\cite{Aaboud:2016vfy,ATL-PHYS-PUB-2015-041}. 
Signal leptons must have $\pT>\SI{20}{\GeV}$. The associated tracks must have a significance of the transverse impact parameter with respect to the reconstructed primary vertex, $d_0$, of $\vert d_0\vert/\sigma(d_0) < 5$ for electrons and $\vert d_0\vert/\sigma(d_0) < 3$ for muons, and a longitudinal impact parameter with respect to the reconstructed primary vertex, $z_0$, satisfying $\vert z_0 \sin\theta\vert  < 0.5$~mm. 
Isolation requirements are applied to both the signal electrons and muons. 
The  scalar sum of the \pt of tracks within a variable-size cone around the lepton, excluding its own track, must be less than 6\% of the lepton \pt. 
The size of the track isolation cone for electrons (muons) is given by the smaller of $\Delta R = \SI{10}{\GeV}/\pt$ and $\Delta R = 0.2\,(0.3)$, 
that is, a cone of size $0.2\,(0.3)$ at low $\pt$ but narrower for high-\pt leptons. 
In addition, in the case of electrons the energy of calorimeter energy clusters in a cone of $\Delta R_\eta = \sqrt{(\Delta\eta)^2+(\Delta\phi)^2} = 0.2$ around the electron 
(excluding the deposition from the electron itself) must be less than 6\% of the electron \pt. 

Simulated events are corrected to account for minor differences in the signal lepton trigger, reconstruction, identification and isolation efficiencies between data and MC simulation.

The missing transverse momentum vector, whose magnitude is denoted by \MET, is defined as the negative vector sum of the transverse momenta of all identified electrons, photons, muons and jets, and an additional soft term. The soft term is constructed from all tracks originating from the primary vertex which are not associated with any identified particle or jet. In this way, the \MET is adjusted for the best calibration of particles and jets listed above, while maintaining pile-up independence in the soft term~\cite{ATL-PHYS-PUB-2015-023,ATL-PHYS-PUB-2015-027}.

The events are classified in a further step into two exclusive categories: at least three leptons plus a $b$-tagged jet (3$\ell$1$b$ selection, aimed at top squark decays involving $Z$ bosons), or at least four $b$-tagged jets and one or two leptons (1$\ell$4$b$ selection, aimed at top squark decays involving Higgs bosons). 

In the 3$\ell$1$b$ selection, events are accepted if they pass a trigger requiring either two electrons, two muons or an electron and a muon. In the 1$\ell$4$b$ selection, events are accepted if they pass a trigger requiring an isolated electron or muon.
The trigger-level requirements on the \pt, identification and isolation of the leptons involved in the trigger decision are looser than those applied offline to ensure that trigger efficiencies are constant in the relevant phase space~\cite{Aaboud:2016leb}. 

Additional requirements are applied depending on the final state, as described in the following. These requirements are optimised for the best discovery significance using the simplified models featuring $\stoptwo$ production with $\stoptwo\to Z\stopone$ or $\stoptwo\to h\stopone$ decays.

\subsection{3$\ell$1$b$ selection}

Events of interest are selected if they contain at least three signal leptons (electrons or muons), with at least one same-flavour opposite-sign lepton pair whose invariant mass is compatible with the $Z$ boson mass ($|m_{\ell\ell}-m_Z|<15$~\GeV, with $m_Z=91.2$~ \GeV). 
To maximise the sensitivity in different regions of the mass parameter space, 
three overlapping signal regions (SRs) are defined as shown in Table~\ref{tab:SRdefZ}.
Signal region \SRAZ\ is optimised for large $\stoptwo - \ninoone$ mass splitting, where the $Z$ boson in the $\stoptwo\to Z\stopone$ decay is boosted, and large $p_{\textrm T}^{\ell\ell}$ and leading-jet $\pt$ are required. Signal region \SRBZ\ covers the intermediate case, featuring slightly softer kinematic requirements than in \SRAZ. Signal region \SRCZ\ is designed to improve the sensitivity for compressed spectra ($m_{\stoptwo} \gtrsim m_{\ninoone} + m_{t} + m_Z$) with softer jet-\pt\ requirements and an upper bound on $p_{\textrm T}^{\ell\ell}$.

\begin{table}[!htb]
\caption{\label{tab:SRdefZ} Definition of the signal regions used in the 3$\ell$1$b$ selection (see text for details).}

\begin{center}
\begin{tabular}{l r r r}
\noalign{\smallskip}\hline\noalign{\smallskip}
Requirement / Region                   & \SRAZ & \SRBZ & \SRCZ \\
\noalign{\smallskip}\hline\noalign{\smallskip}
Number of leptons                      & $\geq 3$ & $\geq 3$ & $\geq 3$   \\
$n_{b-{\textrm{tagged\,jets}}}$                    & $\geq 1$ & $\geq 1$ & $\geq 1$ \\
$|m_{\ell\ell}-m_Z|$ [\GeV]             & $ < 15$  & $ < 15$  & $ < 15$  \\
Leading lepton $p_\text{T}$ [\GeV]      & $>40$    & $>40$    & $>40$    \\
Leading jet $p_\text{T}$ [\GeV]         & $>250$    & $>80$   & $>60$     \\
Leading $b$-tagged jet $p_\text{T}$ [\GeV]     & $>40$    & $>40$    & $>30$    \\
$n_{\textrm jets}$ ($p_{\text{T}}>30$~\GeV) & $\geq 6$ & $\geq 6$ & $\geq 5$ \\
\MET [\GeV]                             & $>100$   & $>180$   & $>140$   \\
$p_{\textrm T}^{\ell\ell}$ [\GeV]           & $>150$   & --       & $<80$    \\
\hline
\end{tabular}
\end{center}
\end{table}

\subsection{1$\ell$4$b$ selection}

Similarly to the 3$\ell$1$b$ case, three overlapping SRs are defined in the 1$\ell$4$b$ selection to have a good sensitivity in different regions of the mass parameter space. 
Only events with one or two signal leptons are selected to ensure orthogonality with the SRs in the 3$\ell$1$b$ selection, with at least one lepton having $\pt>30$~\GeV, and the electron candidates are also required to satisfy the {\textit tight} likelihood-based identification requirement as defined in Refs.~\cite{Aaboud:2016vfy,ATL-PHYS-PUB-2015-041}. 
These SRs are defined as shown in Table~\ref{tab:SRdefH}.

Signal region \SRAH\ is optimised for large $\stoptwo - \ninoone$ mass splitting, where the Higgs boson in the $\stoptwo\to h\stopone$ decay is boosted. In this signal region, the pair of $b$-tagged jets with the smallest $\Delta R^{bb}$ is required to have an invariant mass consistent with the Higgs boson mass ($|m_{bb}-m_h|<15$~\GeV, with $m_h=125$~\GeV), and the transverse momentum of the system formed by these two $b$-tagged jets ($p_{\textrm T}^{bb}$) is required to be above 300~\GeV.
Signal region \SRBH\ covers the intermediate case, featuring slightly harder kinematic requirements than \SRAH.
Finally, signal region \SRCH\ is designed to be sensitive to compressed spectra ($m_{\stoptwo} \gtrsim m_{\ninoone} + m_{t} + m_{h}$). This region has softer jet \pt\ requirements and an upper bound on the $\pt$ of the leading $b$-tagged jet. Signal region \SRAH\ includes requirements on $H_{\textrm T}$ (computed as the scalar sum of the $\pt$ of all the jets in the event), while both signal regions \SRBH\ and \SRCH\ include requirements on the transverse mass $m_{\textrm T}$ computed using the missing-momentum and lepton-momentum vectors: $m_{\textrm T} = \sqrt{2 \pt^{\ell}\met\left(1 - \cos[\Delta\phi(\ell,\met)]\right)}$.

\begin{table}[!htb]
\caption{\label{tab:SRdefH} Definition of the signal regions used in the 1$\ell$4$b$ selection (see text for details).}
\centering
\begin{tabular}{l r r r r}
\noalign{\smallskip}\hline\noalign{\smallskip}
Requirement / Region         & \SRAH    & \SRBH    & \SRCH \\
\noalign{\smallskip}\hline\noalign{\smallskip}
Number of leptons            & 1--2     & 1--2     & 1--2  \\
$n_{b-\textrm{tagged\,jets}}$            & $\geq 4$ & $\geq 4$ & $\geq 4$  \\
$m_{\textrm T}$ [\GeV]            & --       & $>$150   & $>$125 \\
$H_{\textrm T}$ [\GeV]            & $>1000$  & --       & --   \\
\MET [\GeV]                   & $>120$   & $>150$   & $>150$ \\
Leading $b$-tagged jet \pT [\GeV]    & --       & --       & $<$140\\
\mbb [\GeV]                   & 95--155  & --       & --  \\
\ptbb [\GeV]                  & $>300$   & --       & -- \\
$n_{\textrm jets}$ (\pT>60~\GeV) & $\geq 6$ & $\geq 5$ & --  \\
$n_{\textrm jets}$ (\pT>30~\GeV) & --       & --       & $\geq 7$ \\
\hline
\end{tabular}

\end{table}

\FloatBarrier
\section{Background estimation}
\label{sec:bkg}

The main SM background processes satisfying the SR requirements are estimated by simulation, which is normalised and verified (whenever possible) with data events in separate statistically independent regions of the phase space. Dedicated control regions (CRs) enhanced in a particular background component, such as the production of top quark pairs in association with a $Z$ boson ($\ttbar Z$) and multi-boson production in the 3$\ell$1$b$ selection, and $\ttbar$ in the 1$\ell$4$b$ selection, are used for the normalisation. 
For each signal region, a simultaneous ``background fit'' is performed to the numbers of events found in the CRs, using a minimisation based on likelihoods with the HistFitter package \cite{Baak:2014wma}. In each fit, the normalisations of the background contributions having dedicated CRs are allowed to float freely, while the other backgrounds are determined directly using simulation or from additional independent studies in data. This way the total post-fit prediction is forced to be equal to the number of data events in the CR and its total uncertainty is given by the data statistical uncertainty.
When setting 95\% confidence level (CL) upper limits on the cross-section of specific SUSY models, the simultaneous fits also include the observed yields in the SR.

Systematic uncertainties in the MC simulation affect the ratio of the expected yields in the different regions and are taken into account to determine the uncertainty in the background prediction. Each uncertainty source is described by a single nuisance parameter, and correlations between background processes and selections are taken into account. The fit affects neither the uncertainty nor the central value of these nuisance parameters.
The systematic uncertainties considered in the fit are described in Section~\ref{sec:systematics}.

Whenever possible, the level of agreement of the background prediction with data is compared in dedicated validation regions (VRs), which are not used to constrain the background normalisation or nuisance parameters in the fit.

\subsection{Background estimation in the 3$\ell$1$b$ selection}

The dominant SM background contribution to the SRs in the 3$\ell$1$b$ selection is expected to be from $\ttbar Z$, with minor contribution from multi-boson production (mainly $WZ$) and backgrounds containing jets misidentified as leptons (hereafter referred to as ``fake'' leptons) or non-prompt leptons from decays of hadrons (mainly in $t\bar{t}$ events). The normalisation of the main backgrounds ($\ttbar Z$, multi-boson) is obtained by fitting the yield to the observed data in two control regions, then extrapolating this yield to the SRs as described above. Backgrounds from other sources ($\ttbar W$, $\ttbar h$ and rare SM processes), which provide a subdominant contribution to the SRs, are determined from MC simulation only. 

The background from fake or non-prompt leptons is estimated from data with a method similar to that described in Refs.~\cite{top1,top2}. Two types of lepton identification criteria are defined for this evaluation: ``tight'' and ``loose'', corresponding to the signal and candidate electrons and muons described in Section~\ref{sec:selection}. 
The leading lepton is considered to be prompt, which is a valid assumption in more than 95\% of the cases according to simulations. The method makes use of the number of observed events with the second and third leading leptons being loose--loose, loose--tight, tight--loose and tight--tight in each region. The probability for prompt leptons satisfying the loose selection criteria to also satisfy the tight selection is measured using a data sample enriched in $Z\to\ell^+ \ell^-$ ($\ell = e, \mu$) decays. The equivalent probability for fake or non-prompt leptons is measured using events with one electron and one muon with the same charge. The number of events with one or two fake or non-prompt leptons is calculated from these probabilities and the number of observed events with loose and tight leptons.
The modelling of the background from fake or non-prompt leptons is validated in events passing a selection similar to the SRs, but removing the \MET requirements and inverting the $m_{\ell\ell}$ requirements. 

The two dedicated control regions used for the $\ttbar Z$ (\CRTZ) and multi-boson (\CRVV) background estimation in this selection are defined as shown in Table~\ref{tab:CRdef}. 
To ensure orthogonality with the SRs, an upper bound on $\MET <100$~\GeV\ is required in \CRTZ, while a $b$-jet veto is applied in \CRVV.

\begin{table}[!htb]
\caption{\label{tab:CRdef} Definition of the control regions used in the 3$\ell$1$b$ selection. }

\begin{center}
\begin{tabular}{lrr}
\noalign{\smallskip}\hline\noalign{\smallskip}
Requirement / Region   & \CRTZ & \CRVV \\
\noalign{\smallskip}\hline\noalign{\smallskip}
Number of leptons                      & $\geq 3$ & $\geq 3$  \\
$|m_{\ell\ell}-m_Z|$ [\GeV]  & $ < 15$     & $ < 15$      \\
Leading lepton \pT [\GeV]    & $>40$       & $>40$        \\
Leading jet \pT [\GeV]       & $>60$       & $>30$        \\
$n_{b-{\textrm{tagged\,jets}}}$         & $\geq 1$    & 0            \\
$n_{\textrm jets} (\pT>30$~\GeV) & $\geq 4$    & $\geq 4$     \\
\MET [\GeV]                  & $<100$      & --           \\
\(\pt^{\ell\ell}\) [\GeV]    & --          & --           \\
\hline
\end{tabular}
\end{center}
\end{table}

Table~\ref{tab:CRVRZ} shows the observed and expected yields in the two CRs for each background source, and 
Figure~\ref{fig:var_CR} shows the $n_{\textrm jet}$ distribution in these regions after the background fit.
The normalisation factors for the $\ttbar Z$ and multi-boson backgrounds do not differ from unity by more than 30\% 
and the post-fit MC-simulated jet multiplicity distributions agree well with the data.

\begin{figure}[htb!]
\begin{center}
    \begin{subfigure}[t]{0.5\textwidth}
    	\centering
		\includegraphics[width=\textwidth]{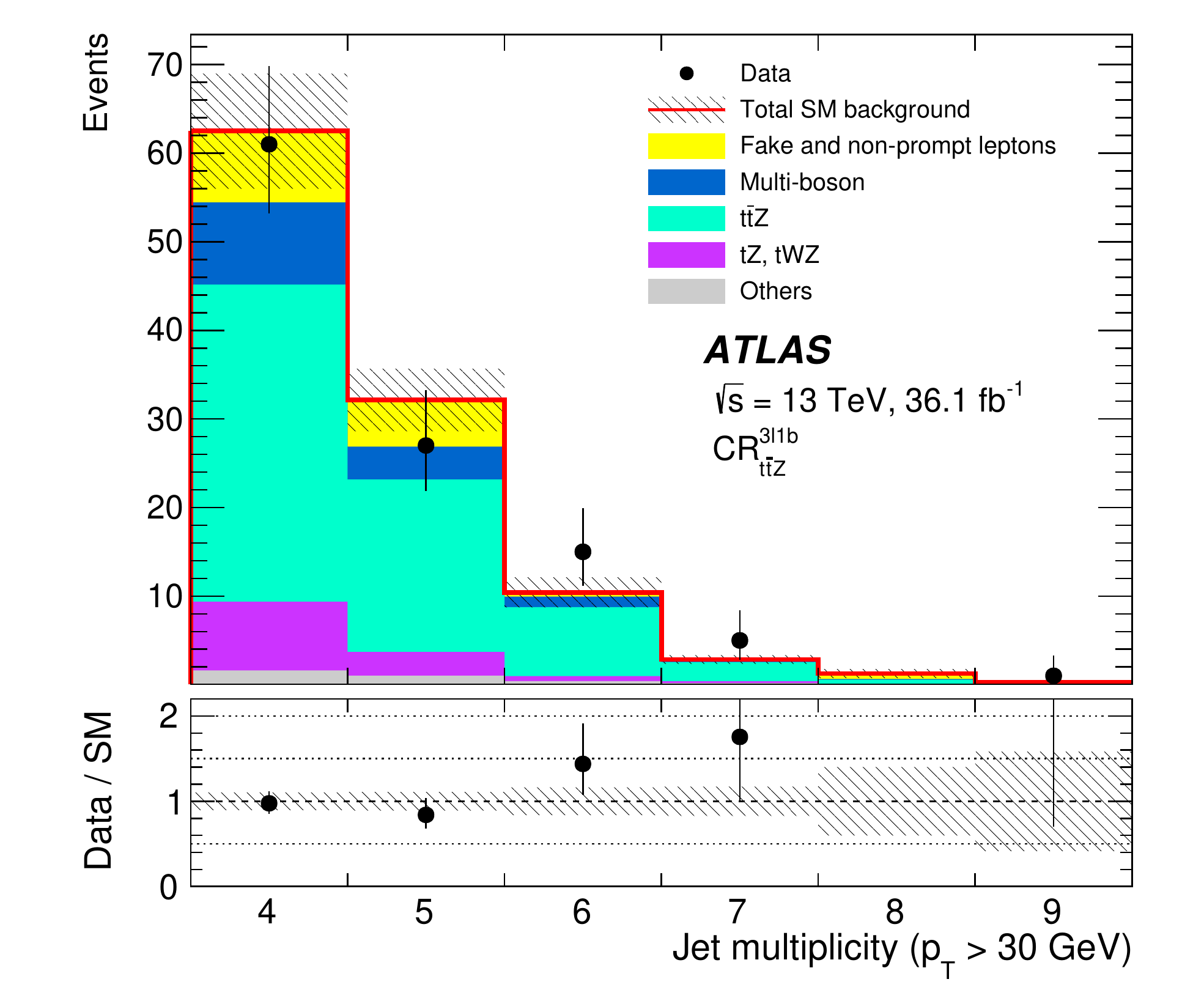}
		\caption{}
	\end{subfigure}%
    \begin{subfigure}[t]{0.5\textwidth}
    	\centering
    	\includegraphics[width=\textwidth]{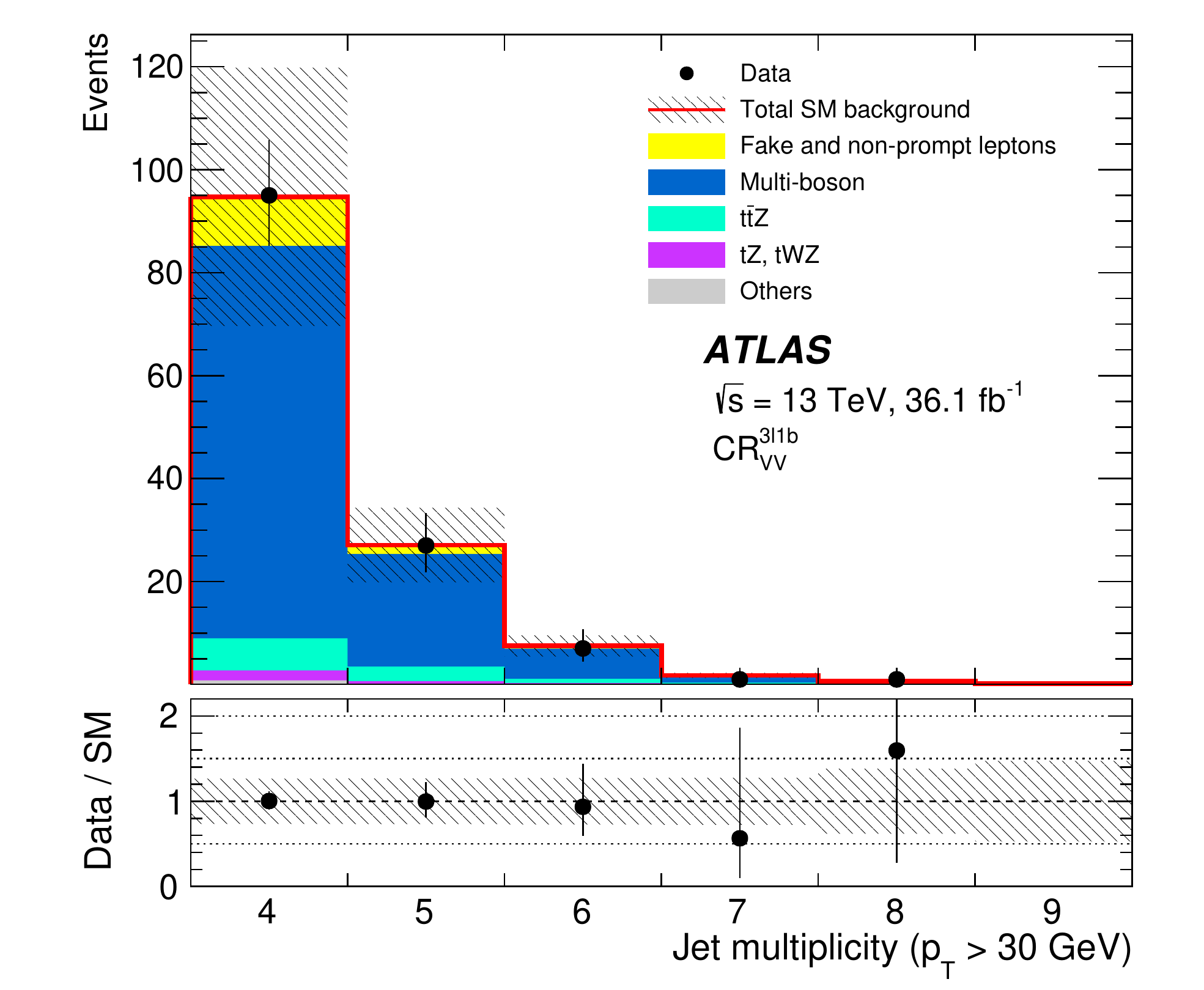}
    	\caption{}
	\end{subfigure}
\end{center}
\caption{Jet multiplicity distributions in control regions (a) \CRTZ\ and (b) \CRVV, after normalising the $\ttZ$ and multi-boson background processes via the simultaneous fit described in Section~\ref{sec:bkg}. The contributions from all SM backgrounds are shown as a histogram stack; the bands represent the total uncertainty in the background prediction. The ``Others'' category contains the contributions from $\ttbar h$, $\ttbar WW$, $\ttbar t$, $\ttbar \ttbar$, $Wh$, and $Zh$ production. 
The last bin in each figure contains the overflow.
The lower panels show the ratio of the observed data to the total SM background prediction, with bands representing the total uncertainty in the background prediction.} 
\label{fig:var_CR}
\end{figure}

\noindent

\begin{table}[!tbh]
\caption{Background fit results for the control regions in the 3$\ell$1$b$ selection. The nominal predictions from MC simulation are given for comparison for those backgrounds ($\ttZ$, multi-boson) that are normalised to data. The ``Others'' category contains the contributions from $\ttbar h$, $\ttbar WW$, $\ttbar t$, $\ttbar \ttbar$, $Wh$, and $Zh$ production. 
Combined statistical and systematic uncertainties are given. The individual uncertainties can be correlated and do not necessarily add in quadrature to the total systematic uncertainty. The number of events with fake or non-prompt leptons is estimated with the data-driven technique described in Section~\ref{sec:bkg}. 
\label{tab:CRVRZ}
}
\begin{center}
\setlength{\tabcolsep}{0.0pc}
{\small
\begin{tabular*}{0.62 \textwidth}{@{\extracolsep{\fill}}lrr}
\noalign{\smallskip}\hline\noalign{\smallskip}
                              & \CRTZ             & \CRVV             \\[-0.05cm]
\noalign{\smallskip}\hline\noalign{\smallskip}
Observed events               & $109$             & $131$                    \\
\noalign{\smallskip}\hline\noalign{\smallskip}
Total (post-fit) SM events & $109 \pm 10$      & $131 \pm 11$              \\
\noalign{\smallskip}\hline\noalign{\smallskip}
Fit output, multi-boson       & $14.5 \pm 2.7$    & $105 \pm 13$              \\
Fit output, \ttZ              & $66 \pm 14$       & $10.2 \pm 2.7$              \\
Fake or non-prompt leptons   & $14 \pm 6$        & $12 \pm 7$              \\
$tZ$, $tWZ$                   & $11 \pm 6$        & $2.7 \pm 1.4$             \\
Others                        & $3.2 \pm 0.5$     & $1.0 \pm 0.3$              \\     
 \noalign{\smallskip}\hline\noalign{\smallskip}
Fit input, multi-boson        & $19$        & $137$              \\
Fit input, \ttZ               & $73$        & $11.2$              \\
\noalign{\smallskip}\hline\noalign{\smallskip}
\end{tabular*}
}
\end{center}
\end{table}

\subsection{Background estimation in the 1$\ell$4$b$ selection}

The dominant SM background contribution to the SRs in the 1$\ell$4$b$ selection is expected to be top quark pair ($\ttbar$) production, amounting to more than 80\% of the total background. 
The normalisation of the $\ttbar$ background for each of the three SRs is obtained by fitting the yield to the observed data in a dedicated CR, then extrapolating this yield to the SRs as described above.
Other background sources (single top, $\ttbar h$ and rare SM processes), which provide a subdominant contribution to the SRs, are determined from MC simulation only. The contribution from events with fake or non-prompt leptons is found to be negligible in this selection.
The three $\ttbar$ CRs (named \CRAH, \CRBH\ and \CRCH) are described in Table~\ref{tab:CRdefH_H}. 
They are designed to have kinematic properties resembling as closely as possible those of each of the three SRs (\SRAH, \SRBH\ and \SRCH, respectively), while having a high purity in $\ttbar$ background and only a small contamination from signal. The CRs are built by inverting the SR requirements on $\met$ and relaxing or inverting those on $m_{bb}$ or $m_{\text T}$. Figure~\ref{fig:var_CRH} shows the jet multiplicity distributions in these CRs after the background fit. In a similar manner, three validation regions (named \VRAH, \VRBH\ and \VRCH) are defined, each of them corresponding to a different CR, with the same requirements on \MET as the SR and relaxing or inverting the requirements on $m_{bb}$, $m_{\text T}$ or jet multiplicity, as shown in Table~\ref{tab:CRdefH_H}. These VRs are used to provide a statistically independent cross-check of the extrapolation in a selection close to that of the SR but with small signal contamination. Table~\ref{tab:CRVR_H} shows the observed and expected yields in the CRs and VRs for each background source. 
The large correction to $\ttbar$ normalisation after the background fit has also been observed in other analyses~\cite{HIGG-2015-05} and is due to a mismodelling of the $\ttbar+\bbbar,\ccbar$ component in the MC simulation.
The background prediction is in agreement with the observed data in all VRs.

\begin{table}[!htb]
\caption{\label{tab:CRdefH_H} Summary of selection criteria for the control and validation regions in the 1$\ell$4$b$ selection.}
\begin{center}
\begin{tabular}{lrrrrrr}
\noalign{\smallskip}\hline\noalign{\smallskip}
Requirement / Region         & \CRAH & \VRAH & \CRBH & \VRBH & \CRCH & \VRCH\\
\noalign{\smallskip}\hline\noalign{\smallskip}
Number of leptons            & 1--2      & 1--2      &  1--2     & 1--2      & 1--2      &  1--2  \\
$n_{b\textrm{-tagged\,jets}}$            & $\geq 4$  & $\geq 4$  & $\geq 4$  & $\geq 4$  & $\geq 4$  & $\geq 4$  \\
$m_{\textrm T}$ [\GeV]            & --        & --        & $>$100    & $>$150    & $<$125    & $<$125 \\
\MET [\GeV]                   & $<$120    & $>$120    & $<150$    & $>$150     & $<150$    & $>$150    \\
Leading $b$-tagged jet \pT [\GeV]    & --        & --        & --        & --         & $<$140    & $<$140 \\
\mbb [\GeV]                   & 95--155   & $\notin$ [95,155]   & --        & --        & --        & --        \\
$n_{\textrm jets}$ (\pT>60~\GeV) & $\geq 5$  & $\geq 5$  & $\geq 5$  & $\leq 4$   & --        & --  \\
$n_{\textrm jets}$ (\pT>30~\GeV) & --        & --        & --        & --         & $\geq 7$  & $\geq 7$ \\
\hline
\end{tabular}
\end{center}
\end{table}

\begin{figure}[htb!]
\begin{center}
    \begin{subfigure}[t]{0.5\textwidth}
    	\centering
		\includegraphics[width=\textwidth]{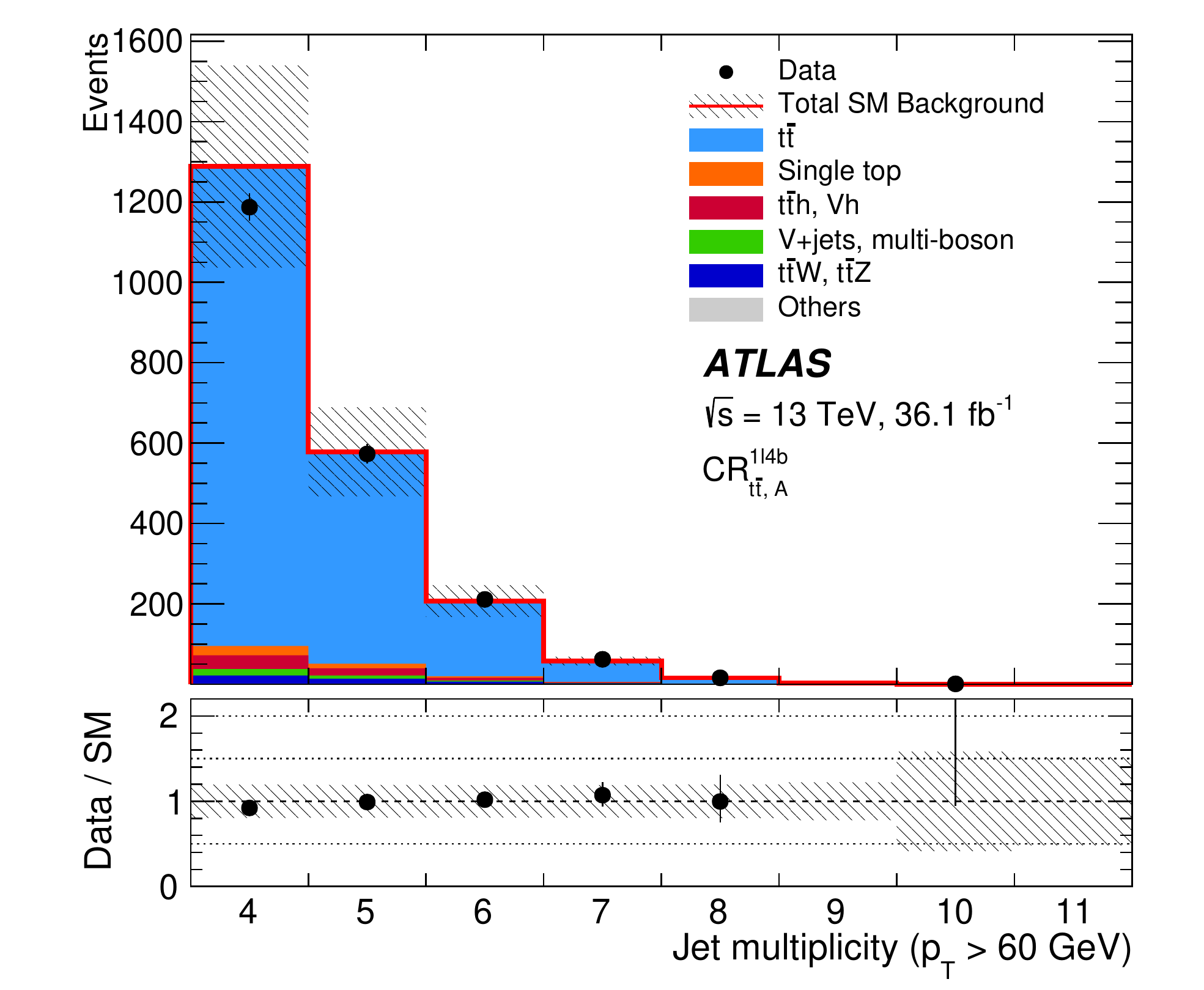}
		\caption{}
	\end{subfigure}%
    \begin{subfigure}[t]{0.5\textwidth}
    	\centering
		\includegraphics[width=\textwidth]{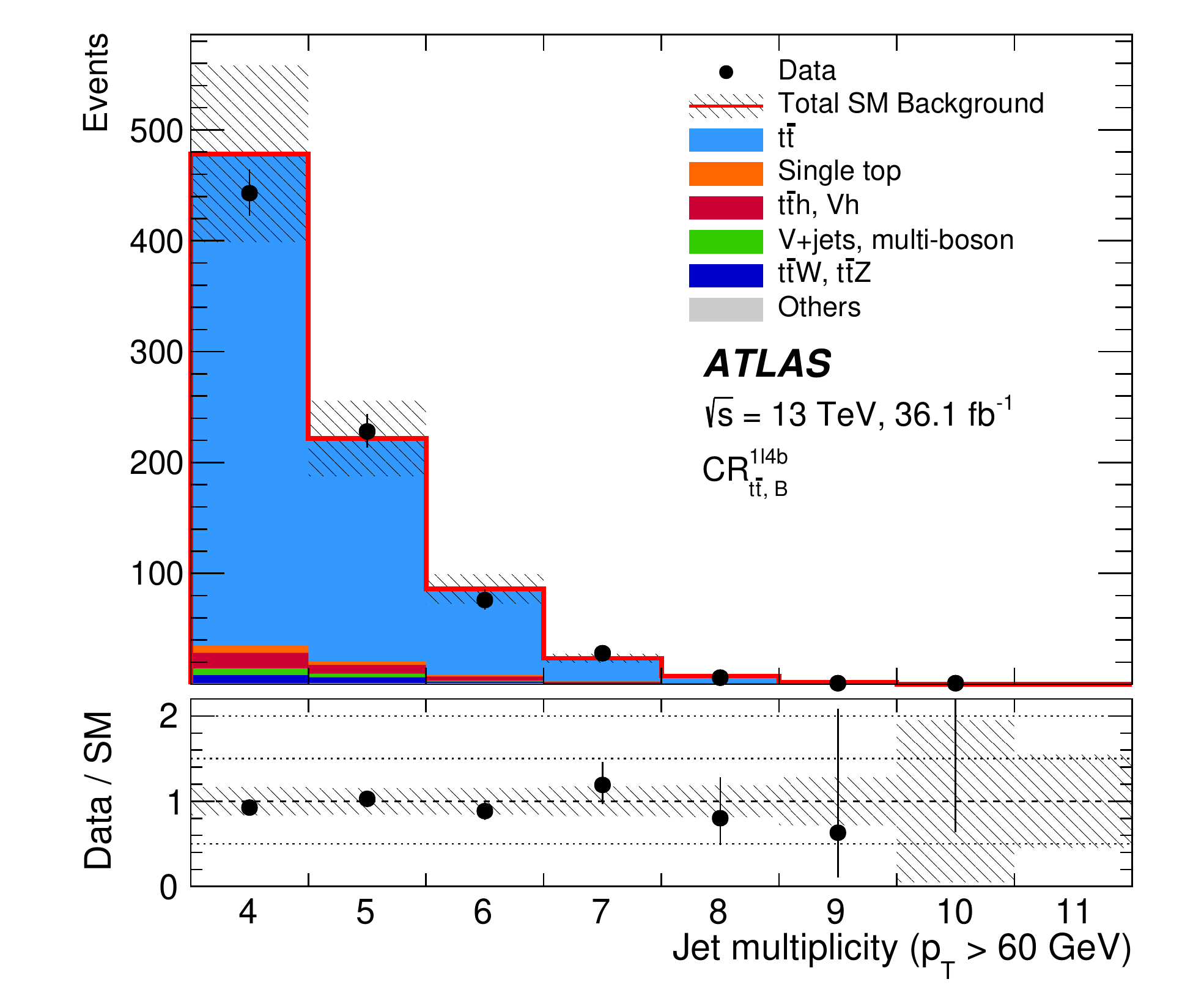}
    	\caption{}
	\end{subfigure} \\
    \begin{subfigure}[b]{0.5\textwidth}
    	\centering
		\includegraphics[width=\textwidth]{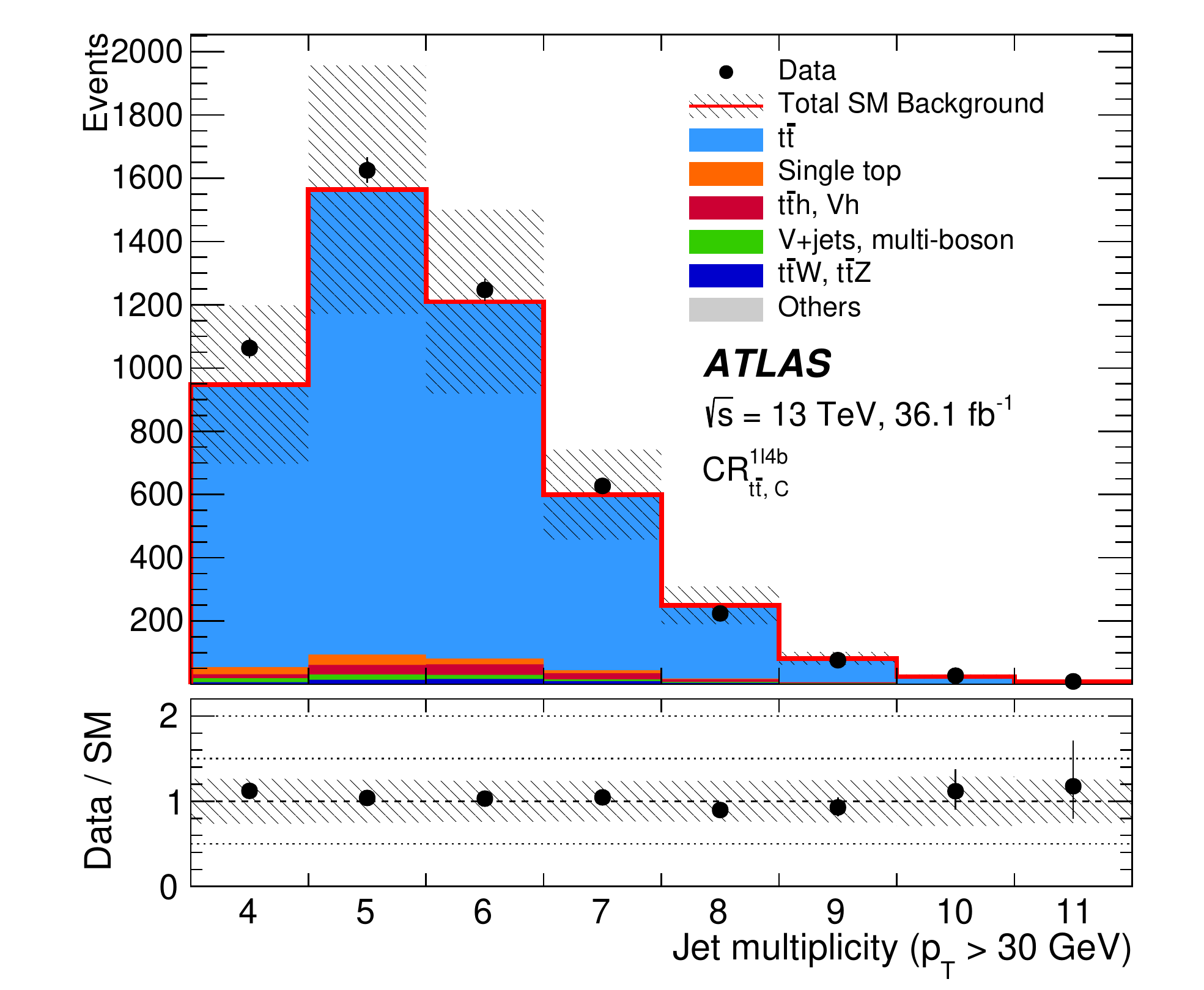}
    	\caption{}
	\end{subfigure}%
\end{center}
\caption{Jet multiplicity distributions in control regions (a) \CRAH\, (b) \CRBH\ and (c) \CRCH\, after normalising the $\ttbar$ background process via the simultaneous fit described in Section~\ref{sec:bkg}. The $\ttbar$ background normalisation is constrained to the data observation for jet multiplicity values above the requirements shown in Table~\ref{tab:CRdefH_H}. The contributions from all SM backgrounds are shown as a histogram stack; the bands represent the total uncertainty. The ``Others'' category contains the contributions from $\ttbar h$, $\ttbar WW$, $\ttbar t$, $\ttbar \ttbar$, $Wh$, and $Zh$ production. The last bin in each figure contains the overflow. The lower panels show the ratio of the observed data to the total SM background prediction, with the bands representing the total uncertainty in the background prediction.} 
\label{fig:var_CRH}
\end{figure}

\begin{table}[!tbh]
\caption{Background fit results for the control and validation regions in the 1$\ell$4$b$ selection. The nominal predictions from MC simulation are given for comparison for the $\ttbar$ background, which is normalised to data. 
The ``Others'' category contains the contributions from $\ttbar WW$, $\ttbar t$, $\ttbar \ttbar$, $tZ$, and $tWZ$ production. 
Combined statistical and systematic uncertainties are given. The individual uncertainties can be correlated and do not necessarily add in quadrature to the total systematic uncertainty. 
\label{tab:CRVR_H}
}
\begin{center}
\setlength{\tabcolsep}{0.0pc}
{\small
\begin{tabular*}{\textwidth}{@{\extracolsep{\fill}}lrrrrrr}
\noalign{\smallskip}\hline\noalign{\smallskip}
                               & \CRAH          & \VRAH         & \CRBH          & \VRBH           & \CRCH          & \VRCH\\[-0.05cm]
\noalign{\smallskip}\hline\noalign{\smallskip}	
Observed events                & $863$          & $258$         & $340$          & $86$            & $963$          & $84$               \\
\noalign{\smallskip}\hline\noalign{\smallskip}
Total (post-fit) SM events  & $863 \pm 29$   & $266 \pm 34$  & $340 \pm 18$   & $96 \pm 13$     & $963 \pm 31$   & $90 \pm 11$        \\
\noalign{\smallskip}\hline\noalign{\smallskip}	
Fit output, \ttbar             & $783 \pm 33$   & $235 \pm 33$  & $307 \pm 19$   & $88 \pm 12$     & $891 \pm 33$   & $82 \pm 10$        \\
Single top                     & $16 \pm 5$     & $9.0 \pm 2.1$ & $5.5 \pm 1.8$  & $1.7 \pm 0.9$   & $12.2 \pm 2.4$ & $2.5 \pm 1.3$      \\
$V$+jets, multi-boson          & $11.8 \pm 2.9$ & $3.1 \pm 1.1$ & $4.7 \pm 1.4$  & $0.15_{-0.15}^{+0.20}$ & $9.8 \pm 2.1$  & $1.0 \pm 0.4$      \\
$\ttbar h$, $Vh$        & $27 \pm 4$     & $7.9 \pm 1.3$ & $12.7 \pm 2.0$ & $3.9 \pm 0.7$   & $31 \pm 5$     & $2.7 \pm 0.6$      \\
\ttW, \ttZ                     & $19 \pm 4$     & $7.2 \pm 1.6$ & $7.1 \pm 1.7$  & $2.2 \pm 0.5$   & $15.6 \pm 3.1$ & $1.5 \pm 0.5$      \\
Others                         & $5.0 \pm 2.6$  & $3.6 \pm 1.9$ & $3.2 \pm 1.7$  & $0.57 \pm 0.31$ & $2.7 \pm 1.4$  & $0.62 \pm 0.32$    \\
\noalign{\smallskip}\hline\noalign{\smallskip}
Fit input, \ttbar              & $495$        & $148$       & $175$        & $50$         & $578$        & $53$             \\
\noalign{\smallskip}\hline\noalign{\smallskip}
\end{tabular*}
}
\end{center}
\end{table}

\FloatBarrier
\section{Systematic Uncertainties}
\label{sec:systematics}

The primary sources of systematic uncertainty are related to the jet energy scale, the jet energy resolution, the theoretical and the MC modelling uncertainties in the background determined using CRs ($\ttbar Z$ and multi-bosons in the 3$\ell$1$b$ selection, as well as $\ttbar$ in the 1$\ell$4$b$ selection). The statistical uncertainty of the simulated event samples is taken into account as well. The effects of the systematic uncertainties are evaluated for all signal samples and background processes. Since the normalisation of the dominant background processes is extracted in dedicated CRs, the systematic uncertainties only affect the extrapolation to the SRs in these cases.

The jet energy scale and resolution uncertainties are derived as a function of the \pT and $\eta$ of the jet, as well as of the pile-up conditions and the jet flavour composition (more quark-like or gluon-like) of the selected jet sample. They are determined using a combination of simulated and data samples, through  measurements of the jet response asymmetry in dijet, $Z$+jet and $\gamma+$jet events \cite{ATL-PHYS-PUB-2015-015}.
Uncertainties associated with the modelling of the $b$-tagging efficiencies for $b$-jets, $c$-jets and light-flavour jets~\cite{ATLAS-CONF-2014-004,ATLAS-CONF-2014-046} are also considered.

The systematic uncertainties related to the modelling of \MET in the simulation are estimated by propagating the uncertainties in the energy and momentum scale of all identified electrons, photons, muons and jets, as well as the uncertainties in the soft-term scale and resolution~\cite{ATL-PHYS-PUB-2015-023}.

Other detector-related systematic uncertainties, such as those in the lepton reconstruction efficiency, energy scale and energy resolution, and in the modelling of the trigger~\cite{PERF-2015-10}, are found to have a small impact on the results.

The uncertainties in the modelling of the $\ttbar$ and single-top backgrounds in simulation in the 1$\ell$4$b$ selection are estimated by varying the renormalisation and factorisation scales, as well as the amount of initial- and final-state radiation used to generate the samples~\cite{ATL-PHYS-PUB-2016-004}. Additional uncertainties in the parton-shower modelling are assessed as the difference between the predictions from \POWHEG showered with \PYTHIA and \HERWIG, and due to the event generator choice by comparing \POWHEG and \AMCATNLO~\cite{ATL-PHYS-PUB-2016-004}, in both cases showered with \PYTHIA.

The diboson background MC modelling uncertainties are estimated by varying the renormalisation, factorisation and resummation scales used to generate the samples~\cite{pubnote_mc_multiboson}.
For $\ttbar Z$, the predictions from the \AMCATNLO and \SHERPA event generators are compared, and the uncertainties related to the choice of renormalisation and factorisation scales are assessed by varying the corresponding event generator parameters up and down by a factor of two around their nominal values~\cite{pubnote_mc_ttv}. 

The cross-sections used to normalise the MC samples are varied according to the uncertainty in the cross-section calculation, i.e., 6\% for diboson, 13\% for $\ttbar W$ and 12\% $\ttbar Z$ production~\cite{Alwall:2014hca}. For $\ttbar WW$, $tZ$, $tWZ$, $\ttbar h$, $\ttbar t$, $\ttbar\ttbar$, and triboson production processes, which constitute a small background, a 50\% uncertainty in the event yields is assumed.

Systematic uncertainties are assigned to the estimated background from fake or non-prompt leptons in the 3$\ell$1$b$ selection to account for potentially different compositions (heavy flavour, light flavour or conversions) between the signal and control regions, as well as the contamination from prompt leptons in the regions used to measure the probabilities for loose fake or non-prompt leptons to pass the tight signal criteria. 

Table~\ref{tab:syst} summarises the contributions of the different sources of systematic uncertainty 
in the total SM background predictions in the signal regions. The dominant systematic uncertainties in the 3$\ell$1$b$ SRs are due to the limited number of events in \CRTZ\ and theoretical uncertainties in $\ttZ$ production, while in the 1$\ell$4$b$ SRs the dominant uncertainties are due to \ttbar modelling.

\begin{table}
\caption{
Summary of the main systematic uncertainties and their impact (in \%) on the
total SM background prediction in each of the signal regions studied. The
total systematic uncertainty can be different from the sum in quadrature of
individual sources due to the correlations between them resulting from the fit
to the data. 
The quoted theoretical uncertainties include modelling and cross-section uncertainties.
 \label{tab:syst}
 }
\begin{center}
\setlength{\tabcolsep}{0.0pc}
{\small
\begin{tabular*}{0.9\textwidth}{@{\extracolsep{\fill}}lrrrrrr}
\noalign{\smallskip}\hline\noalign{\smallskip}
                                  & \SRAZ   & \SRBZ   & \SRCZ & \SRAH & \SRBH & \SRCH \\
  \noalign{\smallskip}\hline\noalign{\smallskip} 
 Total systematic uncertainty (\%)     & 20 & 24 & 15 & 22 & 17 & 30 \\
  \noalign{\smallskip}\hline\noalign{\smallskip}
 Diboson theoretical uncertainties (\%) & 6.7 & 5.5 & 2.2  & <1  & <1  & <1 \\
 \ttZ\ theoretical uncertainties (\%)   & 10  & 10  & 4.4  & <1  & <1  & <1 \\
 \ttbar theoretical uncertainties (\%)  & --    & --    & --     & 17  & 14  & 22 \\
 Other theoretical uncertainties (\%)   & 9.0 & 6.8 & 5.4  & 1.6 & 2.4 & 1.7 \\
  \noalign{\smallskip}\hline\noalign{\smallskip}
 MC statistical uncertainties (\%)      & 8.5 & 18  & 6    & 7.3 & 5.2 & 13\\
  \noalign{\smallskip}\hline\noalign{\smallskip}
 Diboson fitted normalisation (\%)      & 4.6 & 3.5 & 3.8  & <1  & <1  & <1 \\
 \ttZ\ fitted normalisation (\%)        & 12  & 11  & 13   & <1  & <1  & <1 \\
 \ttbar fitted normalisation (\%)       & --    & --    & --     & 3.4 & 5.1 & 3.3 \\
 \noalign{\smallskip}\hline\noalign{\smallskip}
 Fake or non-prompt leptons (\%)        & --    & 6.5 & --     & --    & --    & -- \\
 Pile-up (\%)                           & 4.7 & 2.8 & 0.6 & <1  & 1.4 & <1\\
 Jet energy resolution (\%)             & 2.0 & 2.7 & 3.0  & 5.3 & <1  & 13 \\
 Jet energy scale  (\%)                 & 1.0 & 2.7 & 3.5  & 3.2 & 5.3 & 6.1 \\
 \MET resolution  (\%)                  & 5.3 & 2.6 & 1.6  & 6.8 & 6.5 & 4.0\\
 \(b\)-tagging   (\%)                   & 2.4 & 1.5 & 3.0  & 6.8 & 2.9 & 3.5 \\
\noalign{\smallskip}\hline\noalign{\smallskip}
\end{tabular*}
}
\end{center}

\end{table}

\section{Results}
\label{sec:result}

The observed number of events and expected yields are shown in Table~\ref{tab:results} for each of the six SRs. The SM backgrounds are estimated as described in Section~\ref{sec:bkg}. Data agree with the SM background prediction within uncertainties and thus exclusion limits for several beyond-the-SM (BSM) scenarios are extracted.  
Figure~\ref{fig:MET_SR} shows the $\MET$ distributions after applying all the SR selection requirements except those on $\MET$. 

\begin{figure}[htb!]
\begin{center}

    \begin{subfigure}[t]{0.5\textwidth}
    	\centering
		\includegraphics[width=0.9\textwidth]{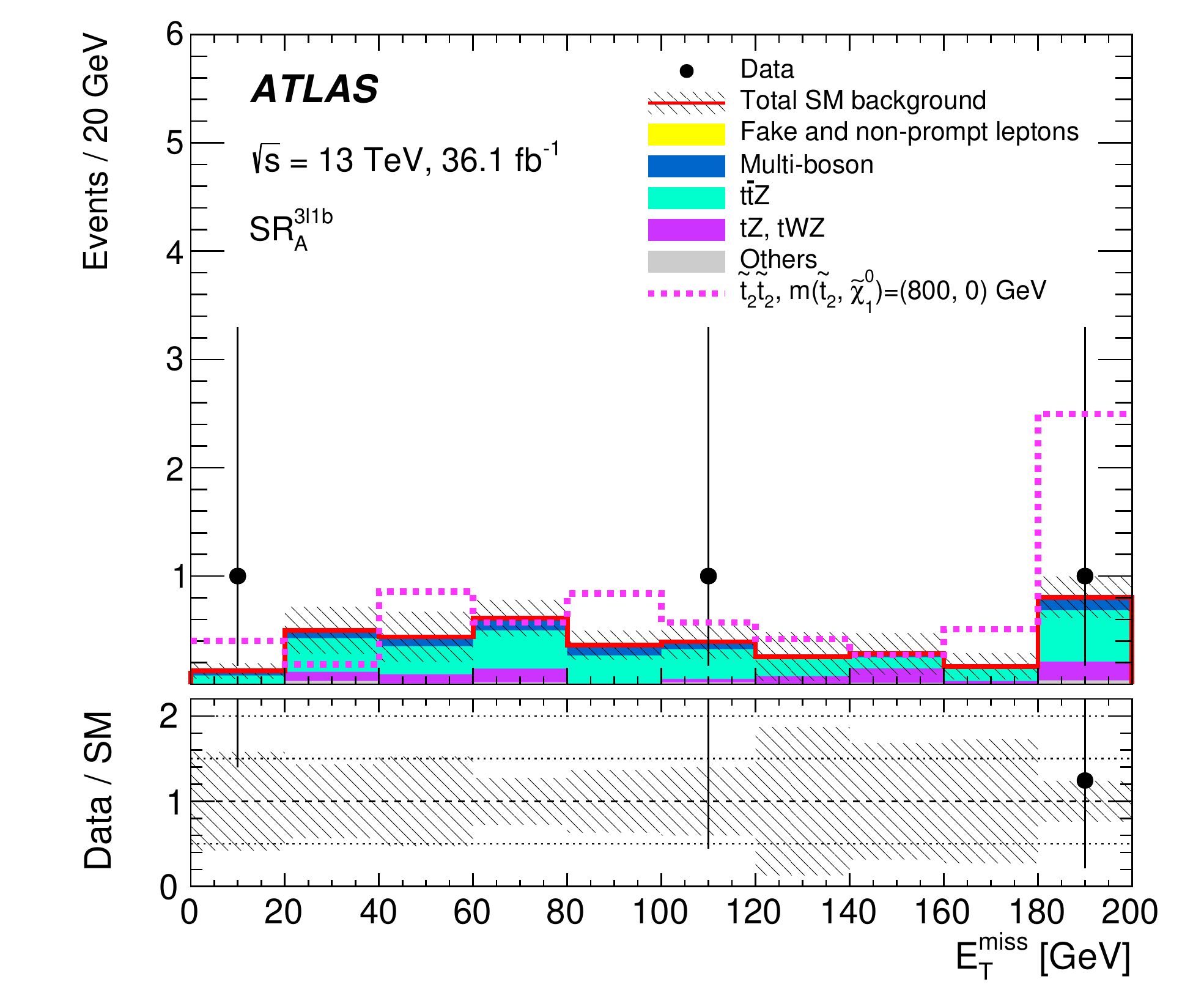}
    	\caption{}
	\end{subfigure}%
    \begin{subfigure}[t]{0.5\textwidth}
    	\centering
		\includegraphics[width=0.9\textwidth]{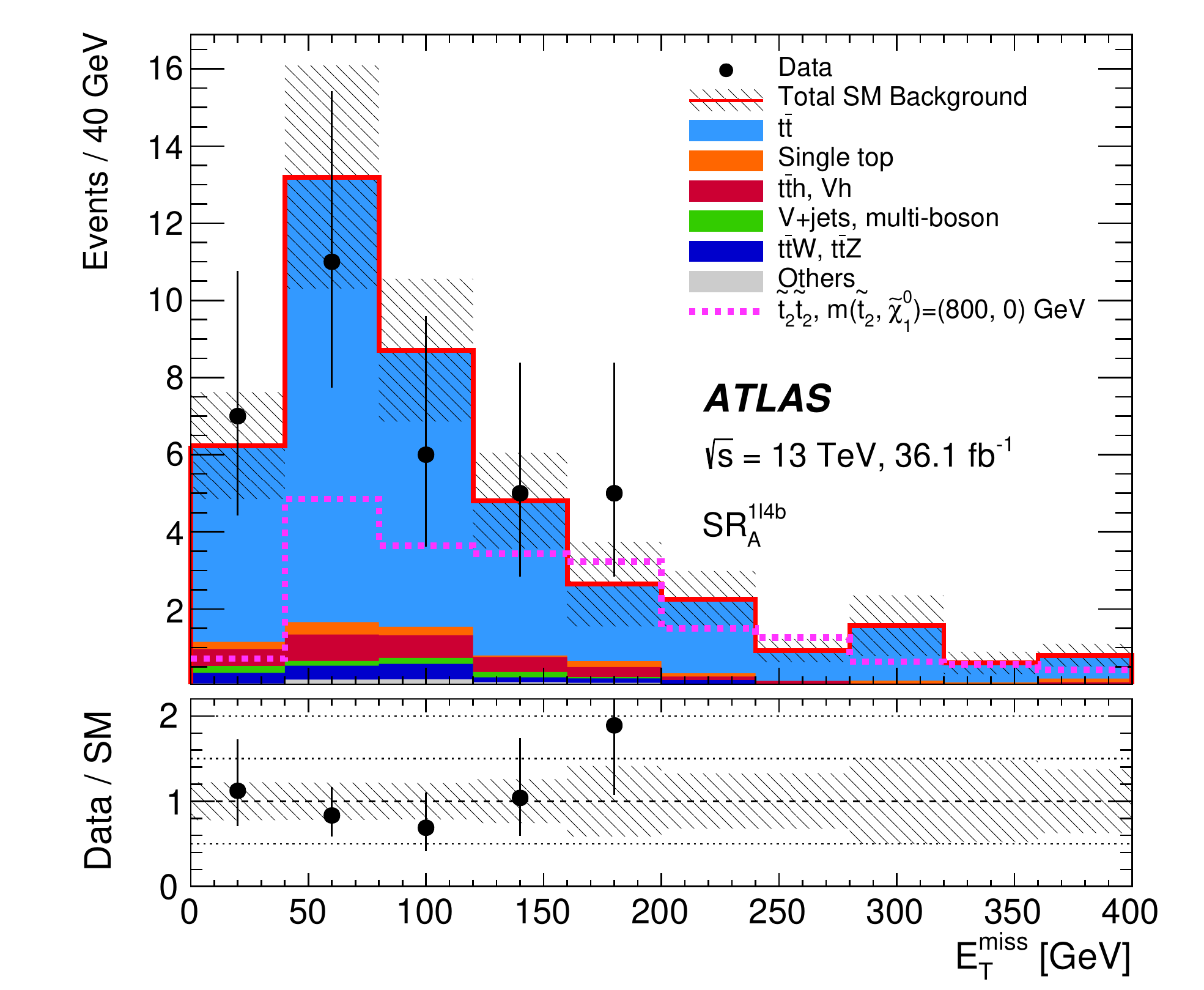}
    	\caption{}
	\end{subfigure} \\
    \begin{subfigure}[t]{0.5\textwidth}
    	\centering
		\includegraphics[width=0.9\textwidth]{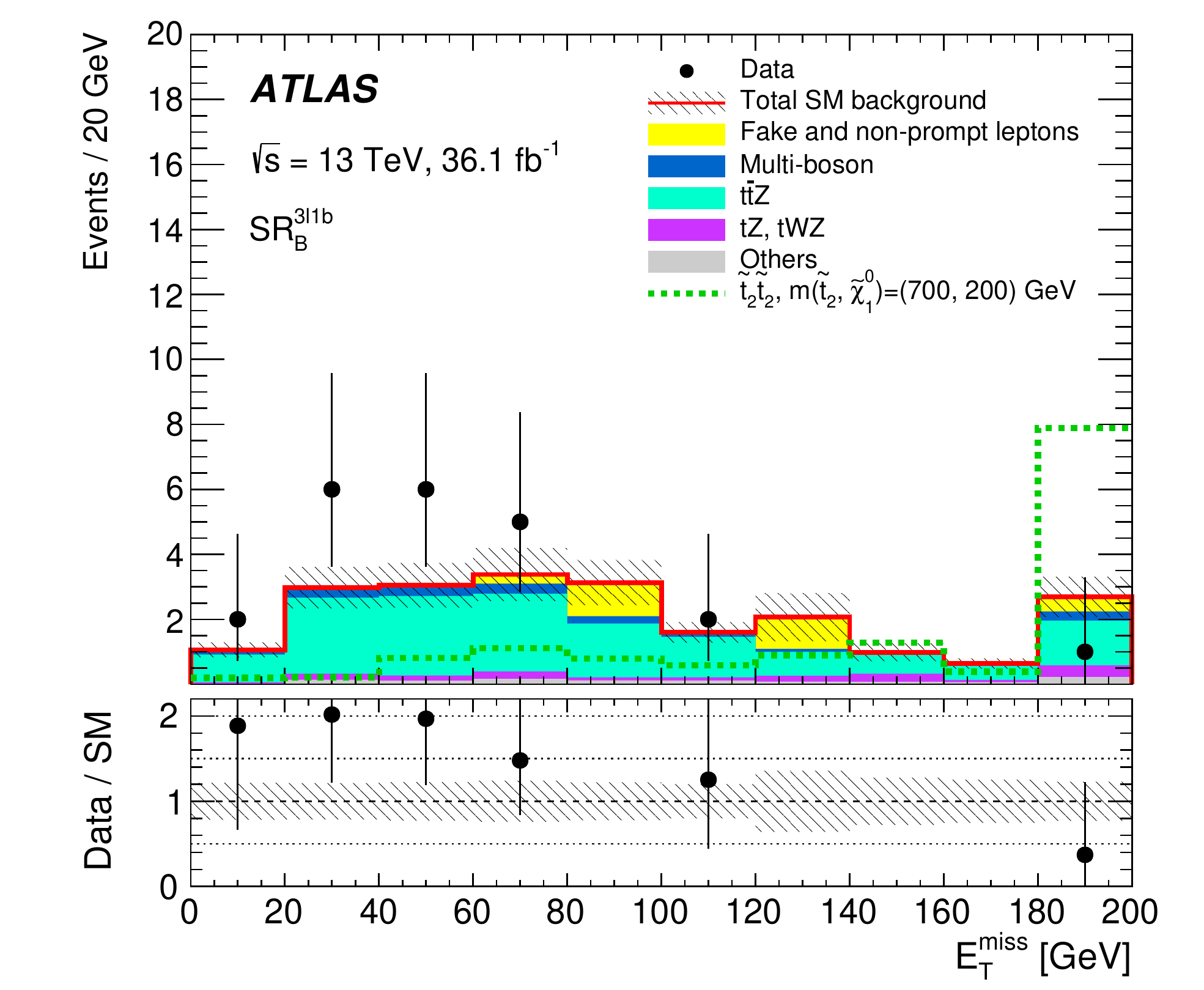}
    	\caption{}
	\end{subfigure}%
    \begin{subfigure}[t]{0.5\textwidth}
    	\centering
		\includegraphics[width=0.9\textwidth]{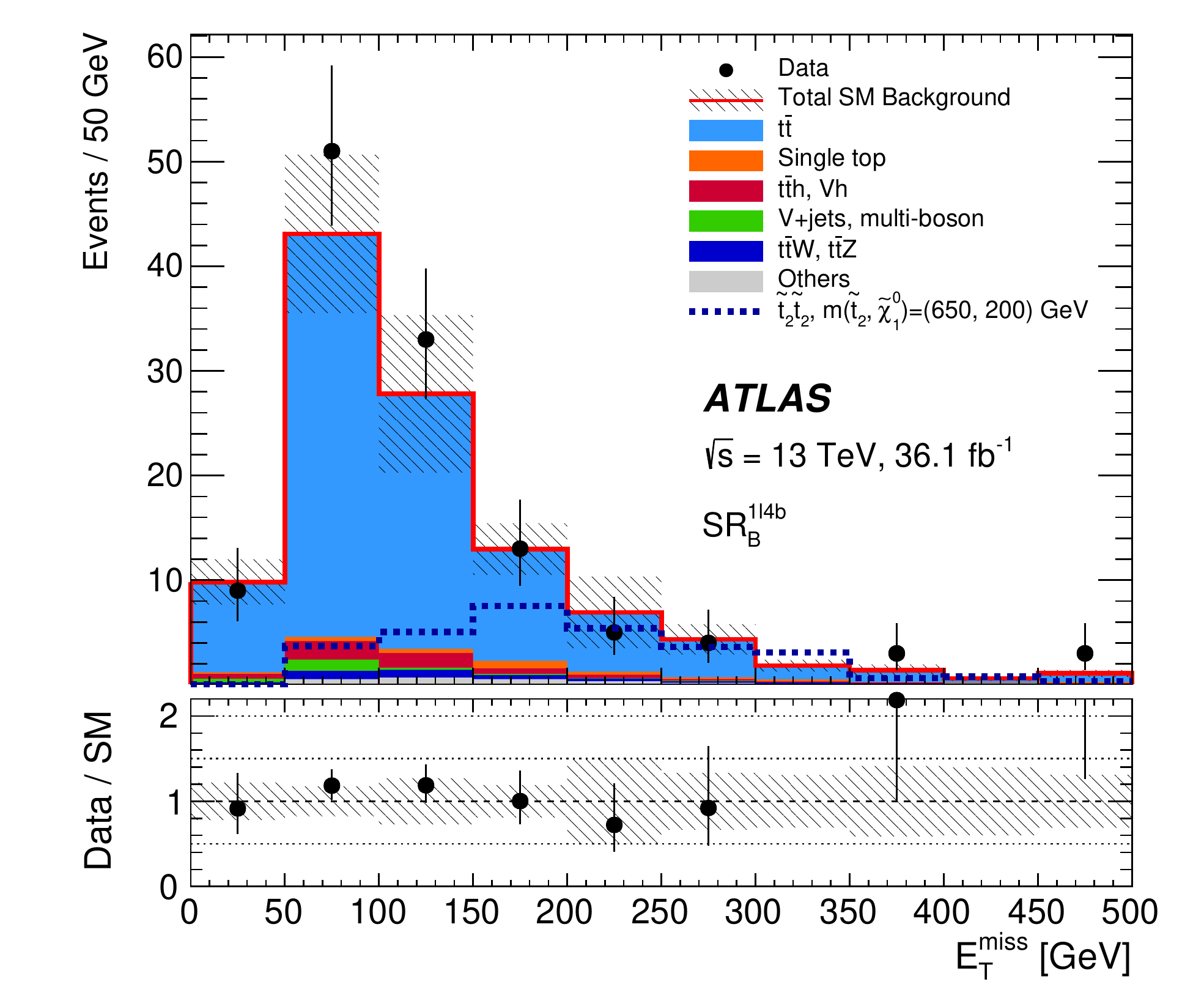}
    	\caption{}
	\end{subfigure} \\
    \begin{subfigure}[t]{0.5\textwidth}
    	\centering
		\includegraphics[width=0.9\textwidth]{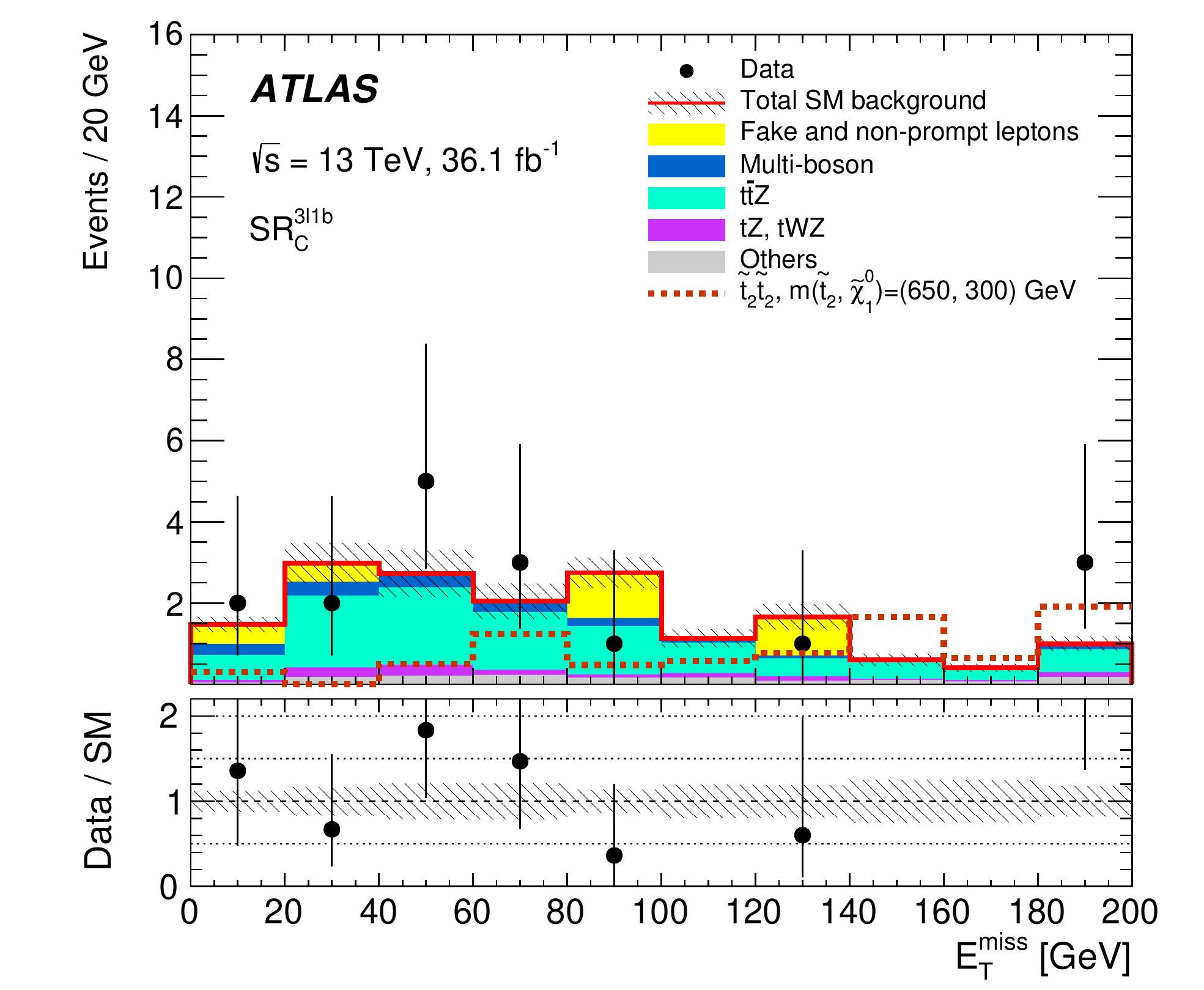}
    	\caption{}
	\end{subfigure}%
    \begin{subfigure}[t]{0.5\textwidth}
    	\centering
		\includegraphics[width=0.9\textwidth]{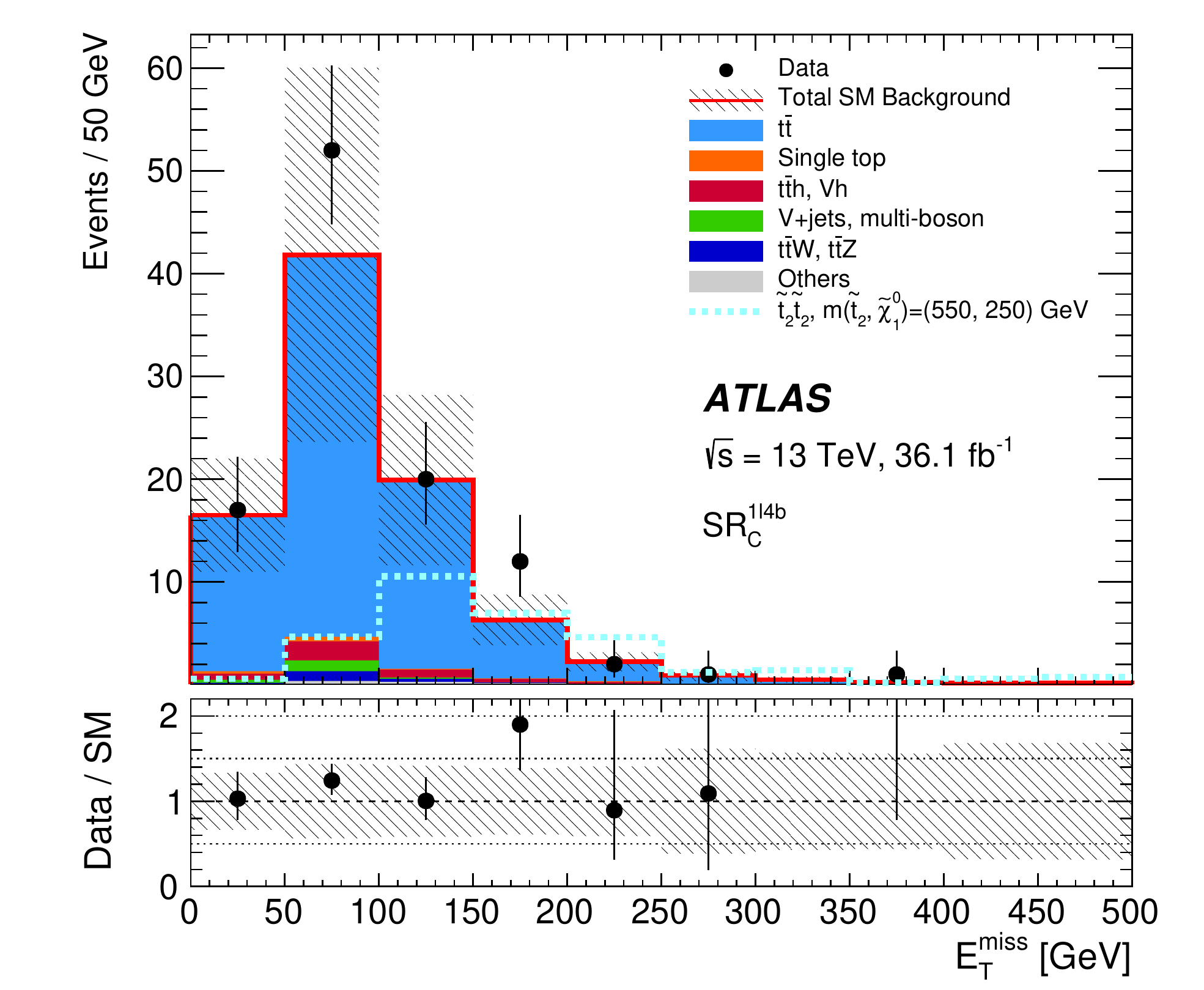}
    	\caption{}
	\end{subfigure}%
\end{center}
\caption{Distribution of \MET\ for events passing all the signal candidate selection requirements, except that on \MET, for (a) \SRAZ, (c) \SRBZ, (e) \SRCZ\, and (b) \SRAH, (d) \SRBH, (f) \SRCH\ after the background fit described in Section~\ref{sec:bkg}. The contributions from all SM backgrounds are shown; the bands represent the total uncertainty. 
The expected distributions for signal models with $m(\stoptwo)=700$~\GeV\ and $m(\ninoone)=0$~\GeV, and $m(\stoptwo)=650$~\GeV\ and $m(\ninoone)=250$~\GeV\ are also shown as dashed lines. 
The last bin in each figure contains the overflow.
The lower panels show the ratio of the observed data to the total SM background prediction, with the bands representing the total uncertainty in the background prediction.}
\label{fig:MET_SR}
\end{figure}

The HistFitter framework, which utilises a profile-likelihood-ratio-test statistic~\cite{Cowan:2010js}, 
is used to estimate 95\% confidence intervals using the CL$_\mathrm{s}$ prescription~\cite{Read_CLs}. 
The likelihood is built as the product of a probability density function describing the observed number of events in the SR and the associated CR(s) and, to constrain the nuisance parameters associated with the systematic uncertainties, Gaussian distributions whose widths correspond to the sizes of these uncertainties;
Poisson distributions are used instead to model statistical uncertainties affecting the observed and predicted yields in the CRs.
Table~\ref{tab:results} also shows upper limits (at the 95\% CL) on the visible BSM cross-section $\sigma_{\mathrm{vis}} = S_{\textrm obs}^{95}/{\cal L}dt$, defined as the product of the production cross-section, acceptance and efficiency. 

Model-dependent limits are also set in specific classes of SUSY models. For each signal hypothesis, the background fit is redone taking into account the signal contamination in the CRs, which is found to be below 15\% for signal models close to the Run-1 exclusion limits. All uncertainties in the SM prediction are considered, including those that are correlated between signal and background (for instance, jet energy scale uncertainties), as well as all uncertainties in the predicted signal, excluding PDF- and scale-induced uncertainties in the theoretical cross-section. Since the three SRs are not orthogonal, only the SR with best expected sensitivity is used for each signal point. ``Observed limits'' are calculated from the observed event yields in the SRs. ``Expected limits'' are calculated by setting the nominal event yield in each SR to the corresponding mean expected background. 

\begin{table}
\caption{Observed and expected numbers of events in the six signal regions. The nominal predictions from MC simulation are given for comparison for those backgrounds ($\ttbar Z$, multi-boson for the 3$\ell$1$b$ selection and \ttbar for the 1$\ell$4$b$ selection) that are normalised to data in dedicated control regions. 
For \SRAZ, \SRBZ\ and \SRCZ, the ``Others'' category contains the contributions from $\ttbar h$, $\ttbar WW$, $\ttbar t$, $\ttbar \ttbar$, $Wh$, and $Zh$ production. 
For \SRAH, \SRBH\ and \SRCH, the ``Others'' category contains the contributions from $\ttbar WW$, $\ttbar t$, $\ttbar \ttbar$, $tZ$, and $tWZ$ production. Combined statistical and systematic uncertainties are given. 
Signal model-independent 95\% CL upper limits on the visible BSM cross-section ($\sigma_{\mathrm{vis}}$), the visible number of signal events ($S^{95}_{\textrm obs}$), the number of signal events ($S^{95}_{\textrm exp}$) given the expected number of background events (and $\pm1\sigma$ variations of the expected background), and the discovery $p$-value ($p(s = 0)$), all calculated with pseudo-experiments, are also shown for each signal region.
 \label{tab:results}
}
\begin{center}
\setlength{\tabcolsep}{0.0pc}
{\small
\def\arraystretch{1.2}
\begin{tabular*}{0.9 \textwidth}{@{\extracolsep{\fill}}lrrr}
\noalign{\smallskip}\hline\noalign{\smallskip}
                                             & \SRAZ            		& \SRBZ                  & \SRCZ                  	\\[-0.05cm]
\noalign{\smallskip}\hline\noalign{\smallskip}
Observed events                              & $2$                  	& $1$                    & $3$                    	\\
\noalign{\smallskip}\hline\noalign{\smallskip}
Total (post-fit) SM events                & $1.9 \pm 0.4$            & $2.7 \pm 0.6$          & $2.0 \pm 0.3$             \\
\noalign{\smallskip}\hline\noalign{\smallskip}
Fit output, multi-boson                      & $0.26 \pm 0.08$          & $0.28 \pm 0.10$        & $0.23 \pm 0.05$           \\
Fit output, \ttZ                             & $1.1 \pm 0.3$            & $1.4 \pm 0.5$          & $1.2 \pm 0.3$             \\
$tZ$, $tWZ$                                  & $0.43 \pm 0.23$          & $0.36 \pm 0.19$        & $0.19 \pm 0.10$           \\
Fake or non-prompt leptons                  & $0.00_{-0.00}^{+0.30}$ 	& $0.45 \pm 0.19$        & $0.00_{-0.00}^{+0.30}$ 	 \\
Others                                       & $0.09 \pm 0.02$          & $0.23 \pm 0.06$        & $0.36 \pm 0.06$           \\
\noalign{\smallskip}\hline\noalign{\smallskip}
Fit input, multi-boson                       & $0.35$              		& $0.37$                 & $0.30$                 	 \\
Fit input, \ttZ                              & $1.2$              		& $1.5$                 & $1.4$                 	 \\
\noalign{\smallskip}\hline\noalign{\smallskip}
$S_{\textrm obs}^{95}$                           & $4.5$ 					& $3.8$                  & $5.8$                  	 \\
$S_{\textrm exp}^{95}$                           & ${4.2}^{+1.9}_{-0.4}$ 	& ${4.9}^{+1.5}_{-1.1}$  & ${4.4}^{+1.8}_{-0.5}$  	 \\
$\sigma_{\mathrm{vis}}$ [fb]                 & 0.13					& 0.10                   & 0.16                   	 \\
$p(s=0)$                                     & 0.42                  & 0.93  & 0.23 \\
\noalign{\smallskip}\hline\noalign{\smallskip}
\noalign{\smallskip}\hline\noalign{\smallskip}
                                             & \SRAH                  & \SRBH                  & \SRCH            \\[-0.05cm]
\noalign{\smallskip}\hline\noalign{\smallskip}                                                                                                             
Observed events                              & $10$                   & $28$                   & $16$ \\
\noalign{\smallskip}\hline\noalign{\smallskip}
Total (post-fit) SM events                & $13.6 \pm 3.0$         & $29 \pm 5$             & $10.5 \pm 3.2$     \\
\noalign{\smallskip}\hline\noalign{\smallskip}                                                              
Fit output, \ttbar                           & $11.3 \pm 2.9$         & $24 \pm 5$             & $9.3 \pm 3.1$      \\
Single top                                   & $0.50 \pm 0.18$        & $1.7 \pm 0.4$          & $0.24 \pm 0.07$    \\
$V$+jets, multi-boson                        & $0.20 \pm 0.15$        & $0.23 \pm 0.10$        & $0.01 \pm 0.01$    \\
$\ttbar h$, $Vh$                      & $0.89 \pm 0.16$        & $1.19 \pm 0.35$        & $0.56 \pm 0.13$    \\
\ttW, \ttZ                                   & $0.36 \pm 0.21$        & $1.09 \pm 0.31$        & $0.10 \pm 0.10$    \\
Others                                       & $0.37 \pm 0.20$        & $1.33 \pm 0.69$        & $0.34 \pm 0.18$    \\
 \noalign{\smallskip}\hline\noalign{\smallskip}                                                             
Fit input, \ttbar                            & $7.1$                  & $14$                 & $6.0$              \\
\noalign{\smallskip}\hline\noalign{\smallskip}   
$S_{\textrm obs}^{95}$                           & $7.8$                 & $14.6$         & $15.6$ \\
$S_{\textrm exp}^{95}$                           & ${9.6}^{+4.1}_{-2.3}$ & ${15.5}^{+5.6}_{-4.4}$  & ${10.4}^{+4.2}_{-2.6}$  \\
$\sigma_{\mathrm{vis}}$ [fb]                 & 0.21                  & 0.40 & 0.43 \\ 
$p(s=0)$                                     & 0.63                  & 0.82 & 0.11 \\
\noalign{\smallskip}\hline\noalign{\smallskip}
\end{tabular*}
}
\end{center}
\end{table}

Figure~\ref{fig:limstop1} shows the limits on simplified models in which pair-produced \stopone decay with 100\% branching ratio into the \ninotwo and a top quark, assuming ${\cal B}(\ninotwo \rightarrow Z\ninoone) = 0.5$ and ${\cal B}(\ninotwo \rightarrow h\ninoone) = 0.5$. A massless LSP and a minimum mass difference between the \ninotwo and \ninoone of 130~\GeV, needed to have on-shell decays for both the Higgs and \Zboson\ bosons, are assumed in this model. Limits are presented in the \stopone--\ninotwo mass plane. The two SRs with best expected sensitivity from the 3$\ell$1$b$ and 1$\ell$4$b$ selections are statistically combined to derive the limits on this model. For a \ninotwo mass above 200~\GeV, \stopone masses up to about 800~\GeV\ are excluded at 95\% CL.

\begin{figure}
\begin{center}
\includegraphics[width=0.8 \textwidth]{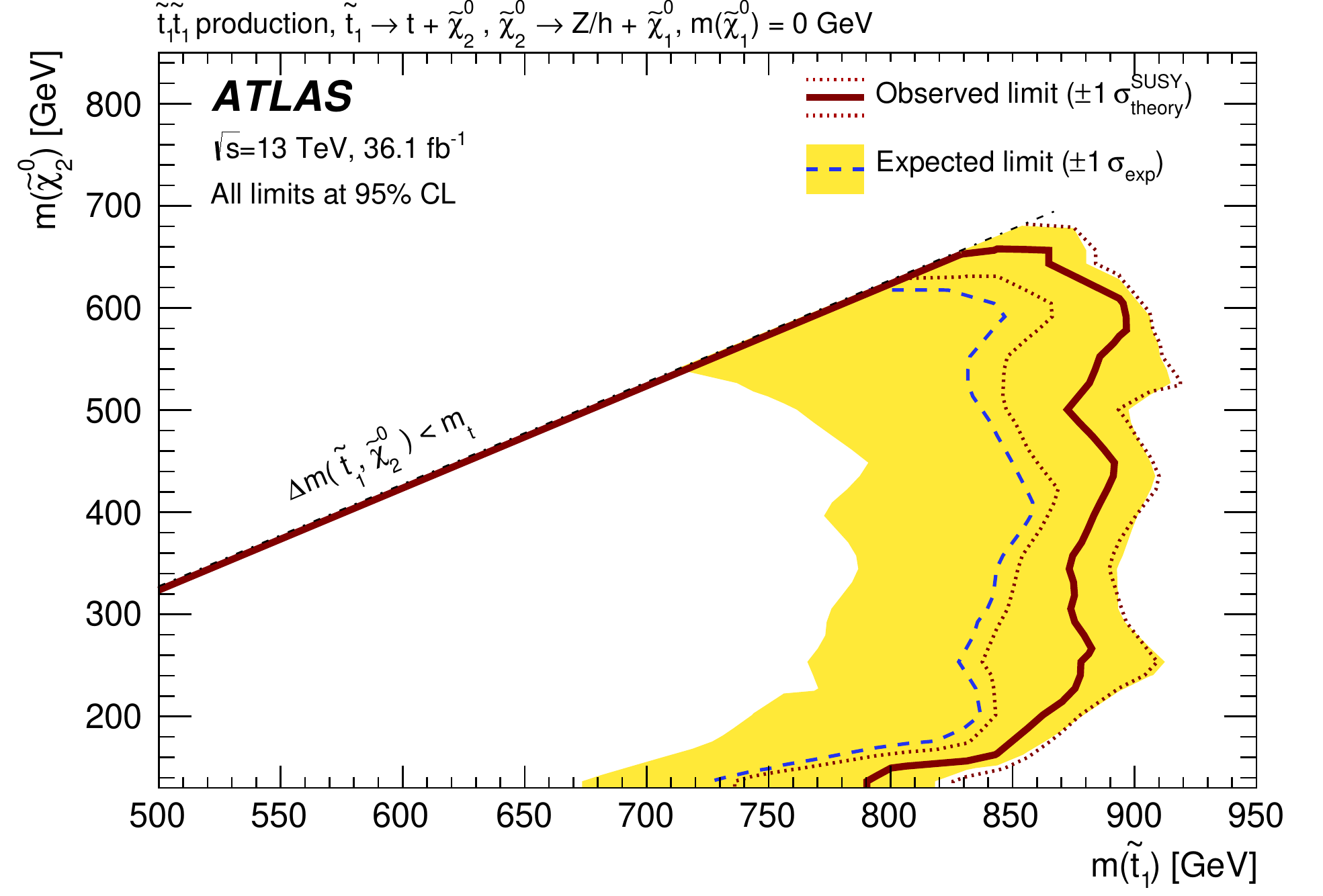}
\end{center}
\caption{Exclusion limits at 95\% CL from the analysis of $36.1 \; \ifb$ of 13~\TeV\ $pp$ collision data on the masses of the $\stopone$ and $\ninotwo$, for a fixed $m(\ninoone) = 0$~\GeV, assuming ${\cal B}(\ninotwo \rightarrow Z\ninoone) = 0.5$ and ${\cal B}(\ninotwo \rightarrow h\ninoone) = 0.5$. 
The dashed line and the shaded band are the expected limit and its $\pm1 \sigma$ uncertainty, respectively. The thick solid line 
is the observed limit for the central value of the signal cross-section.
The expected and observed limits do not include the effect of the theoretical uncertainties in the signal cross-section. The dotted lines show the effect on the observed limit when varying the signal cross-section by $\pm 1 \sigma$ of the theoretical uncertainty. }
\label{fig:limstop1}
\end{figure}

Limits for simplified models, in which pair-produced \stoptwo decay with 100\% branching ratio into the \stopone and either a $Z$ or a $h$ boson, with $\stopone \to t\ninoone$, in the \stoptwo--\ninoone mass plane are shown in Figure~\ref{fig:limstop2}.
When considering the \stoptwo decays via a \Zboson boson, probed by the $3\ell 1b$ selection, \stoptwo masses up to 800~\GeV\ are excluded at 95\% CL for a \ninoone of about 50~\GeV\, and \ninoone masses up to 350~\GeV\ are excluded for \stoptwo masses below 650~\GeV. Assuming 100\% branching ratio into \stopone and a $h$ boson, probed by the $1\ell 4b$ selection, \stoptwo masses up to 880~\GeV\ are excluded at 95\% CL for a \ninoone of about 50~\GeV, and \ninoone masses up to 260~\GeV\ are excluded for \stoptwo masses between 650 and 710~\GeV. These results extend the previous limits on the \stoptwo mass from ATLAS $\sqrt{s}=8$~\TeV\ analyses~\cite{AtlasSummary,Stop2_Run1} by up to 250~\GeV\ depending on the \ninoone mass.

\begin{figure}
\begin{center}
    \begin{subfigure}[t]{0.85\textwidth}
    	\centering
		\includegraphics[width=0.95\textwidth]{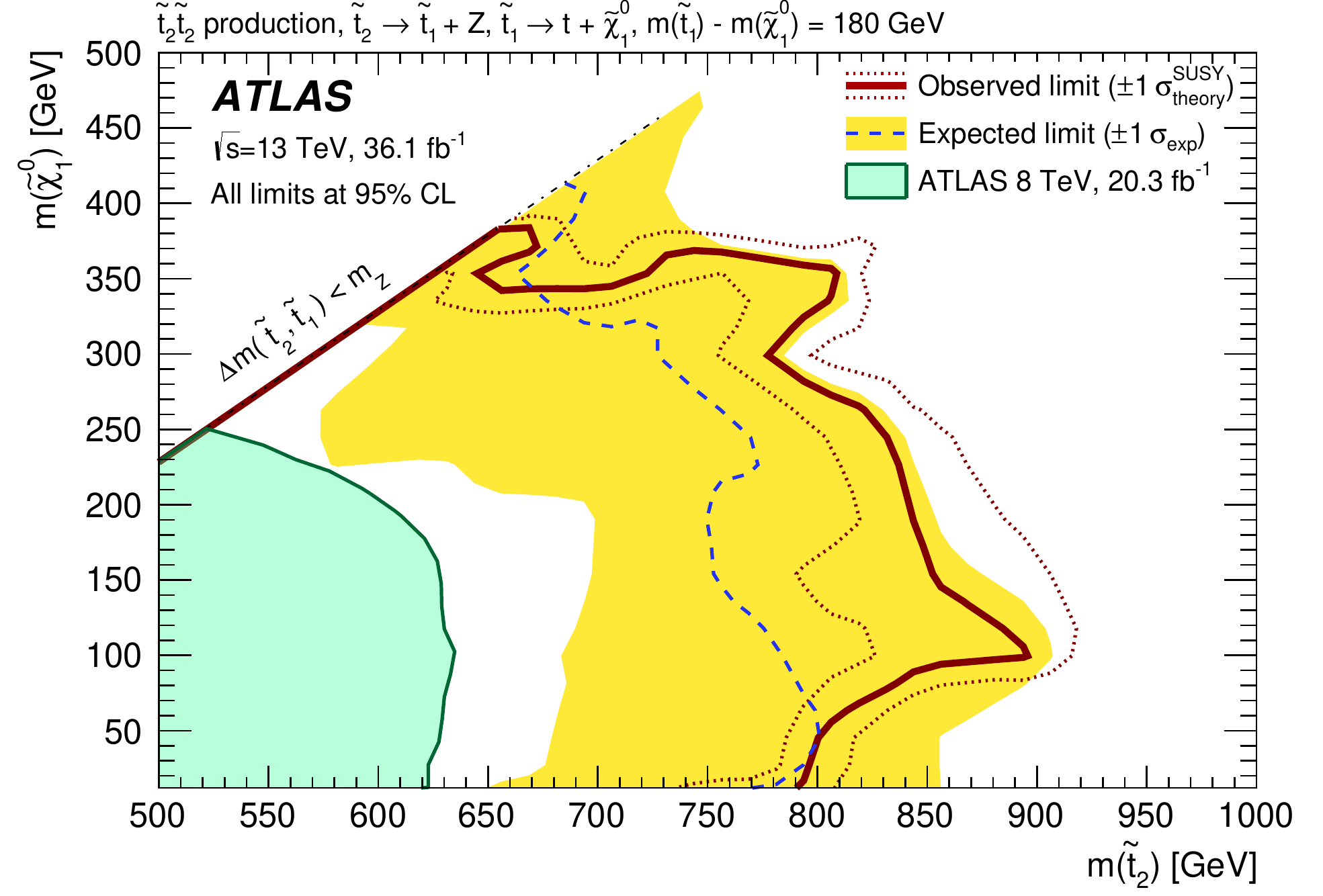}
    	\caption{}
	\end{subfigure}\\\vspace{0.5cm}
    \begin{subfigure}[t]{0.85\textwidth}
    	\centering
		\includegraphics[width=0.95\textwidth]{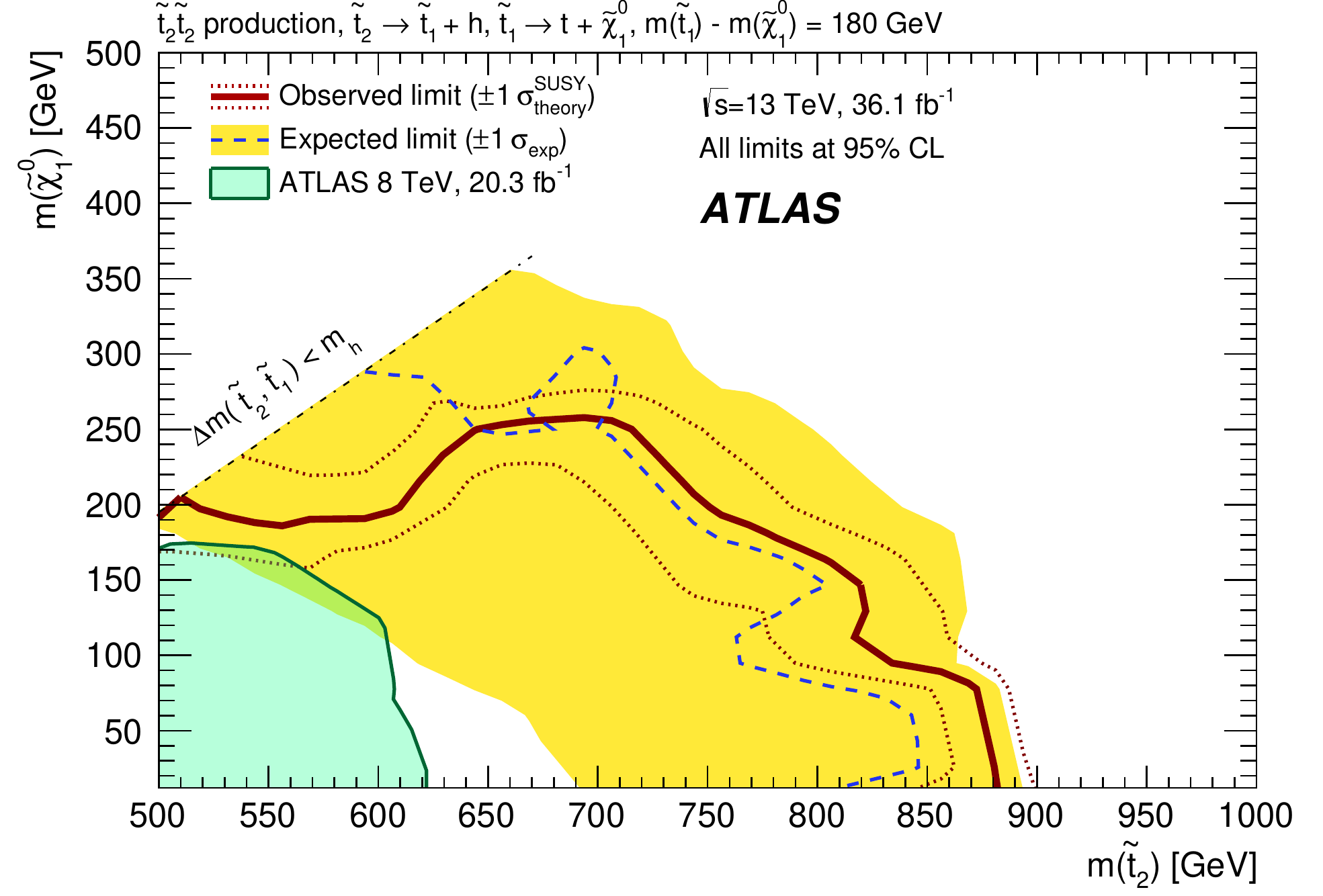}
    	\caption{}
	\end{subfigure} 

\end{center}
\caption{Exclusion limits at 95\% CL from the analysis of $36.1 \; \ifb$ of 13~\TeV\ $pp$ collision data on the masses of the $\stoptwo$ and $\ninoone$, for a fixed $m(\stopone)-m(\ninoone) = 180$~\GeV\ and assuming (a) ${\cal B}(\stoptwo \rightarrow Z\stopone) = 1$ or (b) ${\cal B}(\stoptwo \rightarrow h\stopone) = 1$. The dashed line and the shaded band are the expected limit and its $\pm1 \sigma$ uncertainty, respectively. The thick solid line is the observed limit for the central value of the signal cross-section. The expected and observed limits do not include the effect of the theoretical uncertainties in the signal cross-section. The dotted lines show the effect on the observed limit when varying the signal cross-section by $\pm 1 \sigma$ of the theoretical uncertainty. The shaded area in the lower-left corner shows the observed exclusion from the ATLAS $\sqrt{s}=8$~\TeV\ analysis~\cite{AtlasSummary,Stop2_Run1}.}
\label{fig:limstop2}
\end{figure}

Exclusion limits as a function of the $\stoptwo$ branching ratios are shown in Figure~\ref{fig:limtern} for representative values of the masses of $\stoptwo$ and \ninoone. 
For $\stoptwo$ mass of 600~\GeV, SUSY models with ${\cal B}(\stoptwo\to Z\stopone)$ above 58\% are excluded. For higher top squark mass ($m_{\stoptwo}=650$~\GeV), models with ${\cal B}(\stoptwo\to Z\stopone)$ above 50\% or ${\cal B}(\stoptwo\to h\stopone)$ above 80\% are excluded. The region with large ${\cal B}(\stoptwo\to Z\ninoone)$ can be probed by searches targeting  direct $\stopone$ pair production~\cite{AtlasSummary}.

\begin{figure}
\begin{center}
\includegraphics[width=\textwidth]{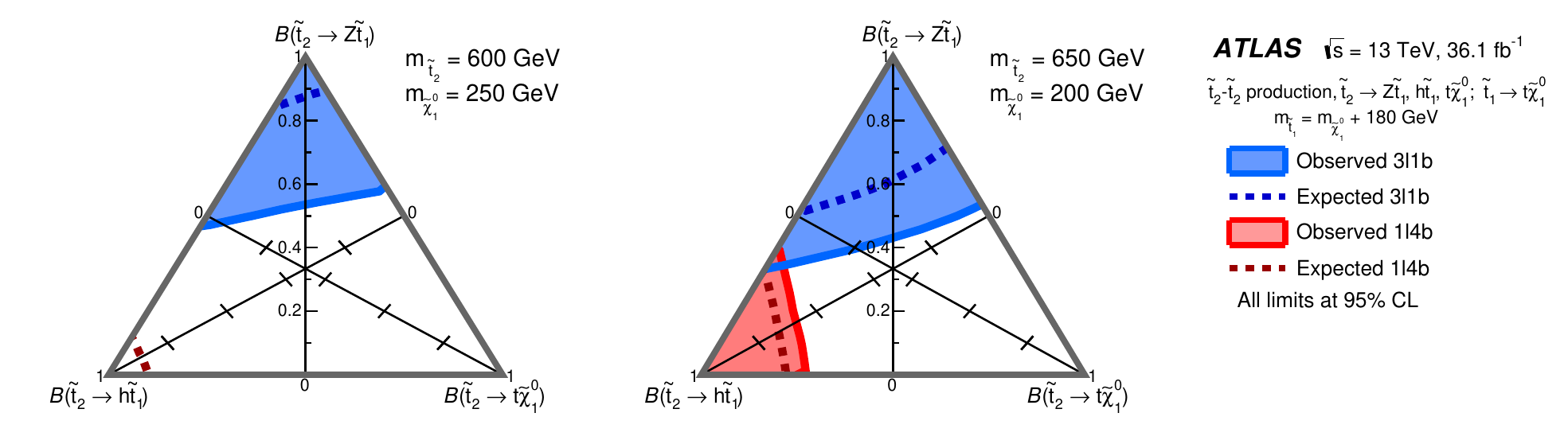}
\end{center}
\caption{Exclusion limits at 95\% CL from the analysis of $36.1 \; \ifb$ of 13~\TeV\ $pp$ collision data as a function of the \stoptwo branching ratio for $\stoptwo \to \stopone\Zboson$, $\stoptwo \to \stopone h$ and $\stoptwo \to t\ninoone$. The blue and red exclusion regions correspond to the 3$\ell$1$b$ and 1$\ell$4$b$ selections respectively. The limits are given for two different values of the \stoptwo and \ninoone masses. The dashed lines are the expected limit and the solid lines are the observed limit for the central value of the signal cross-section.
}
\label{fig:limtern}
\end{figure}


\FloatBarrier

\section{Conclusion}
\label{sec:conclusion}

This paper reports a search for direct top squark pair production resulting in events with either a leptonically decaying $Z$ boson or a pair of $b$-tagged jets from a Higgs boson decay, based on 36.1~\ifb\ of proton--proton collisions at $\sqrt{s} = 13$~\TeV\ recorded by the ATLAS experiment at the LHC in 2015 and 2016. Good agreement is found between the yield of observed events and the SM predictions.
Model-independent limits are presented, which allow the results to be reinterpreted in generic models that also predict similar final states in association with invisible particles. The limits exclude, at 95\% confidence level, beyond-the-SM processes with visible cross-sections above 0.11 (0.21)~fb for the 3$\ell$1$b$ (1$\ell$4$b$) selections.

Results are also interpreted in the context of simplified models characterised by the decay chain $\stopone \to t\ninotwo$ with $\ninotwo \to Z/h\ninoone$, or $\stoptwo\to Z/h\stopone$ with $\stopone\to t\ninoone$.
The results exclude at 95\% confidence level $\tilde{t}_2$ and $\tilde{t}_1$ masses up to about 800~\GeV, extending the region of supersymmetric parameter space covered by previous LHC searches.

\section*{Acknowledgements}


We thank CERN for the very successful operation of the LHC, as well as the
support staff from our institutions without whom ATLAS could not be
operated efficiently.

We acknowledge the support of ANPCyT, Argentina; YerPhI, Armenia; ARC, Australia; BMWFW and FWF, Austria; ANAS, Azerbaijan; SSTC, Belarus; CNPq and FAPESP, Brazil; NSERC, NRC and CFI, Canada; CERN; CONICYT, Chile; CAS, MOST and NSFC, China; COLCIENCIAS, Colombia; MSMT CR, MPO CR and VSC CR, Czech Republic; DNRF and DNSRC, Denmark; IN2P3-CNRS, CEA-DSM/IRFU, France; SRNSF, Georgia; BMBF, HGF, and MPG, Germany; GSRT, Greece; RGC, Hong Kong SAR, China; ISF, I-CORE and Benoziyo Center, Israel; INFN, Italy; MEXT and JSPS, Japan; CNRST, Morocco; NWO, Netherlands; RCN, Norway; MNiSW and NCN, Poland; FCT, Portugal; MNE/IFA, Romania; MES of Russia and NRC KI, Russian Federation; JINR; MESTD, Serbia; MSSR, Slovakia; ARRS and MIZ\v{S}, Slovenia; DST/NRF, South Africa; MINECO, Spain; SRC and Wallenberg Foundation, Sweden; SERI, SNSF and Cantons of Bern and Geneva, Switzerland; MOST, Taiwan; TAEK, Turkey; STFC, United Kingdom; DOE and NSF, United States of America. In addition, individual groups and members have received support from BCKDF, the Canada Council, CANARIE, CRC, Compute Canada, FQRNT, and the Ontario Innovation Trust, Canada; EPLANET, ERC, ERDF, FP7, Horizon 2020 and Marie Sk{\l}odowska-Curie Actions, European Union; Investissements d'Avenir Labex and Idex, ANR, R{\'e}gion Auvergne and Fondation Partager le Savoir, France; DFG and AvH Foundation, Germany; Herakleitos, Thales and Aristeia programmes co-financed by EU-ESF and the Greek NSRF; BSF, GIF and Minerva, Israel; BRF, Norway; CERCA Programme Generalitat de Catalunya, Generalitat Valenciana, Spain; the Royal Society and Leverhulme Trust, United Kingdom.

The crucial computing support from all WLCG partners is acknowledged gratefully, in particular from CERN, the ATLAS Tier-1 facilities at TRIUMF (Canada), NDGF (Denmark, Norway, Sweden), CC-IN2P3 (France), KIT/GridKA (Germany), INFN-CNAF (Italy), NL-T1 (Netherlands), PIC (Spain), ASGC (Taiwan), RAL (UK) and BNL (USA), the Tier-2 facilities worldwide and large non-WLCG resource providers. Major contributors of computing resources are listed in Ref.~\cite{ATL-GEN-PUB-2016-002}.

\clearpage

\printbibliography

\newpage
\begin{flushleft}
{\Large The ATLAS Collaboration}

\bigskip

M.~Aaboud$^\textrm{\scriptsize 137d}$,
G.~Aad$^\textrm{\scriptsize 88}$,
B.~Abbott$^\textrm{\scriptsize 115}$,
O.~Abdinov$^\textrm{\scriptsize 12}$$^{,*}$,
B.~Abeloos$^\textrm{\scriptsize 119}$,
S.H.~Abidi$^\textrm{\scriptsize 161}$,
O.S.~AbouZeid$^\textrm{\scriptsize 139}$,
N.L.~Abraham$^\textrm{\scriptsize 151}$,
H.~Abramowicz$^\textrm{\scriptsize 155}$,
H.~Abreu$^\textrm{\scriptsize 154}$
% ...
\bigskip
\\
% ...
$^{12}$ Institute of Physics, Azerbaijan Academy of Sciences, Baku, Azerbaijan\\
% ...
$^{88}$ CPPM, Aix-Marseille Universit{\'e} and CNRS/IN2P3, Marseille, France\\
% ...
$^{115}$ Homer L. Dodge Department of Physics and Astronomy, University of Oklahoma, Norman OK, United States of America\\
% ...
$^{119}$ LAL, Univ. Paris-Sud, CNRS/IN2P3, Universit{\'e} Paris-Saclay, Orsay, France\\
% ...
$^{137}$ $^{(a)}$ Facult{\'e} des Sciences Ain Chock, R{\'e}seau Universitaire de Physique des Hautes Energies - Universit{\'e} Hassan II, Casablanca; $^{(b)}$ Centre National de l'Energie des Sciences Techniques Nucleaires, Rabat; $^{(c)}$ Facult{\'e} des Sciences Semlalia, Universit{\'e} Cadi Ayyad, LPHEA-Marrakech; $^{(d)}$ Facult{\'e} des Sciences, Universit{\'e} Mohamed Premier and LPTPM, Oujda; $^{(e)}$ Facult{\'e} des sciences, Universit{\'e} Mohammed V, Rabat, Morocco\\
% ...
$^{139}$ Santa Cruz Institute for Particle Physics, University of California Santa Cruz, Santa Cruz CA, United States of America\\
% ...
$^{151}$ Department of Physics and Astronomy, University of Sussex, Brighton, United Kingdom\\
% ...
$^{154}$ Department of Physics, Technion: Israel Institute of Technology, Haifa, Israel\\
$^{155}$ Raymond and Beverly Sackler School of Physics and Astronomy, Tel Aviv University, Tel Aviv, Israel\\
% ...
$^{161}$ Department of Physics, University of Toronto, Toronto ON, Canada\\
% ...
$^{a}$ Also at Department of Physics, King's College London, London, United Kingdom\\
$^{b}$ Also at Institute of Physics, Azerbaijan Academy of Sciences, Baku, Azerbaijan\\
$^{c}$ Also at Novosibirsk State University, Novosibirsk, Russia\\
$^{d}$ Also at TRIUMF, Vancouver BC, Canada\\
$^{e}$ Also at Department of Physics {\&} Astronomy, University of Louisville, Louisville, KY, United States of America\\
% ...
$^{*}$ Deceased
\end{flushleft}


\end{document}